\documentclass[%
 preprint,
 amsmath,amssymb,
 aps,
 prb,
]{revtex4-2}

\usepackage{graphicx}
\usepackage{bm}
\def\mb{\mu_{\rm B}}
\def\tc{T_{\rm C}}
\def\ha{H_{\rm a}}
\def\js{J_{\rm s}}

\def\ku{K_{\rm u}}
\def\yfeco{{\rm YFe_3Co_2}}
\def\ycuco{{\rm YCu_3Co_2}}
\def\ynico{{\rm YCo_3Ni_2}}
\def\ycuni{{\rm YCu_3Ni_2}}
\def\yfeni{{\rm YFe_3Ni_2}}
\def\yc{{\rm YCo_5}}
\def\yco{{\yc}}
\def\ycofecuni{{\rm Y(}{\rm Co}_{1-x-y}{\rm Fe}_x{\rm Cu}_y)_3({\rm Co}_{1-z}{\rm Ni}_z)_2}

\def\ycofeni{{\rm Y(}{\rm Co}_{1-x}{\rm Fe}_x)_3({\rm Co}_{1-z}{\rm Ni}_z)_2}
\def\ycocuni{{\rm Y(}{\rm Co}_{1-y}{\rm Cu}_y)_3({\rm Co}_{1-z}{\rm Ni}_z)_2}
\def\yconi{{\rm Y}{\rm Co}_3({\rm Co}_{1-z}{\rm Ni}_z)_2}
\def\ycocu{{\rm Y(}{\rm Co}_{1-y}{\rm Cu}_y)_3{\rm Co}_2}
\def\ycofe{{\rm Y(}{\rm Co}_{1-x}{\rm Fe}_x)_3{\rm Co}_2}

\def\mjmc{{\rm MJ/m^3}}
\def\mjm{{\rm MJ/m^3}}
\def\mjmt{{\rm MJ/m^3}}

\newcommand{\resec}[1]{\mbox{Sec. \ref{#1}}}
\newcommand{\retable}[1]{\mbox {Table \ref{#1}}}
\newcommand{\refig}[1]{\mbox{Fig. \ref{#1}}}

\begin{document}


\title{First-principles Calculation of Magnetocrystalline Anisotropy of Y(Co,Fe,Ni,Cu)$_5$ Based on Full-potential KKR Green's Function Method}

\author{Haruki Okumura}
\affiliation{Institute for Solid State Physics, The University of Tokyo, Kashiwa 277-8581, Japan}

\author{Tetsuya Fukushima}%
\affiliation{Institute for Solid State Physics, The University of Tokyo, Kashiwa 277-8581, Japan}
\affiliation{Institute for AI and Beyond, The University of Tokyo, Bunkyo-ku, Tokyo 113-8656, Japan}

\author{Hisazumi Akai}%
\affiliation{Institute for Solid State Physics, The University of Tokyo, Kashiwa 277-8581, Japan}

\author{Masako Ogura}%
\affiliation{Department of Chemistry and Physical Chemistry, LMU Munich, Butenandtstrasse 11, D-81377 Munich, Germany}

\date{\today}

\begin{abstract}
The performance of permanent magnets $\yc$ can be improved by replacing cobalt with other elements, such as iron, copper, and nickel.
In order to determine its optimum composition, it is necessary to perform systematic theoretical calculations in a consistent framework.
In this study, we calculated the magnetocrystalline anisotropy constant $\ku$ of $\ycofecuni$ on the basis of the full-potential Korringa-Kohn-Rostoker Green's function method in conjunction with the coherent potential approximation.
The calculated $\ku$ of $\yc$ was smaller than the experimental value because of a missing enhancement due to orbital polarization.
Although the value of $\ku$ of $\ycofecuni$ was systematically underestimated compared to their experimental counterparts, the doping effect can be analyzed within a consistent framework.
The results have shown that $\yfeco$ has much higher $\ku=5.00$ $\mjmc$ than pristine YCo$_5$ ($\ku=1.82$ $\mjmc$), and that nickel as a stabilization element decreases $\ku$ and magnetization in YFe$_3$(Co$_{1-z}$Ni$_z$)$_2$. However, the anisotropy field of $z\sim0.5$ can compete with the value of $\yc$.
\end{abstract}

\maketitle

\section{Introduction}
Permanent magnets have been widely used in many applications, including magnetic memory devices, wind power generators, electric vehicles, etc.
As permanent magnets are almost always used above room temperature, their Curie temperature ($\tc$) should be as high as possible.
In addition, high magnetization ($\js$) and high uniaxial magnetocrystalline anisotropy constants ($\ku$) are also essential for high-performance permanent magnets.
Ferromagnetic transition metal elements like iron and cobalt usually exhibit high $\tc$ and $\js$. Large axial anisotropy originates from rare-earth elements with a strong spin-orbit coupling of $4f$-electrons.

One of the most extensively used permanent magnets is Nd-Fe-B alloy, which has high uniaxial anisotropy and coercivity \cite{Sagawa1984, Croat1984}; however, its limitation is low $\tc$ without the addition of dysprosium (an expensive element) (Fig.4 in Ref.\cite{Coey2020}). Another permanent magnet used is SmCo$_5$, which has high $\tc\sim 1000$ K and $\ku\sim 17.2$ $\mjm$ \cite{Tatsumoto1971,Coey2011}. Although its magnetic properties do not decay even at high temperatures, its low-temperature magnetization is smaller than that of Nd$_2$Fe$_{14}$B and Sm$_2$Fe$_{17}$N$_3$, as can be seen by the stoichiometric composition ratios of the transition metals and rare-earths.
Although cobalt is more expensive than iron, various studies investigate improving the performance of SmCo$_5$ because samarium is less costly than neodymium and has high-temperature tolerance.

$\yc$, which has the same crystal structure as SmCo$_5$, has moderate performance as a magnet with experimental values of $\ku\sim 6.5$ $\mjm$, $\tc=978$ K and magnetic moment of $8.3$ $\mb$/f.u. \cite{Tatsumoto1971,Alameda1981,Klein1974,Klein1975}.
The ferromagnetism and large anisotropy are mainly due to cobalt rather than yttrium, which does not have $4f$-electrons.
Since yttrium is less expensive than neodymium, $\yc$ magnets are preferred for mass production.
In general, when the anisotropy comes mainly from the $3d$-electron of the cobalt sublattice instead of $4f$-electrons, the influence of temperature on the magnetic properties is reduced.
Magnetic interactions between cobalt atoms are more significant than $3d$-$4f$ interactions.
In alloys with $4f$-electrons, the anisotropy rapidly decreases as temperature increases because of their weak magnetic interactions \cite{Zhao1991,Skomski1998}.
In $\yc$, however, the experiment indicated that the anisotropy hardly changes in the vicinity of the room temperature \cite{Tatsumoto1971,Klein1974}.

In the past decades, the anisotropy and magnetic properties of pristine $\yc$ have been theoretically calculated \cite{Nordstorm1992,Daalderop1996,Yamaguchi1996,Zhu2014,Sakurai2018,Nguyen2018,Matsumoto2015,Steinbeck2001,Larson2003,Larson2004,Liu2010,Patrick2017,Patrick2019,Asali2019}.
Some researchers replaced a fraction of cobalt with iron and/or copper to enhance the axial anisotropy of $\yc$, \cite{Steinbeck2001,Larson2003,Larson2004,Liu2010,Patrick2017,Patrick2019,Asali2019}, and others investigated the magnetic properties of YCo$_{5-x}$Ni$_x$ \cite{Crisan1995,Yamada1999,Ishikawa2003,Patrick2017,Landa2020}.
Landa $et$ $al.$ presented that proposed YFe$_3$(Co$_{1-z}$Ni$_z$)$_2$ has higher energy products than $\yc$,
where iron increases the magnetization and nickel thermodynamically stabilizes the 1-5 phases \cite{Landa2020}.
Many theoretical calculations for both pristine and doped $\yc$ are available; however, there are few systematic studies within a consistent framework.

In this study, we have systematically calculated $\ku$ and $\js$ of $\ycofecuni$ on the basis of the full-potential Korringa-Kohn-Rostoker (FPKKR) Green's function method \cite{Ogura2005}. The anisotropic potential is more realistic than spherical potentials, such as muffin-tin potential or the potential obtained by the atomic sphere approximation.
The potential in disordered alloys is treated in the framework of the coherent potential approximation (CPA) \cite{Shiba1971,Soven1970}, within which the atomic sites of impurities are thought to be ideally randomized.

\section{Method}
The FPKKR method is based on Green's function formulation, where the multiple-scattering problem is solved \cite{Ogura2005}.
The potential includes a non-spherical part. The space is divided into Voronoi cells, and the CPA \cite{Shiba1971,Soven1970} is employed to consider the disorder in the potentials.
In contrast to the supercell method, the CPA enables the addition of arbitrary concentrations of impurities to the host material without enlarging the unit cell.

The exchange-correlation functionals are chosen as the local density approximation (LDA) parameterized by Moruzzi-Janak-Williams \cite{Moruzzi1978} and the generalized gradient approximation (GGA) parameterized by Perdew-Burke-Ernzerhof \cite{Perdew1996}.
Spin-orbit interaction is included as $l_zs_z$ and the orbital polarization enhancement is not considered.
In this study, therefore, the orbital moment of cobalt atoms could be smaller than experimental values \cite{Heidemann1975,Schweizer1980} or other calculations with orbital polarization correction \cite{Steinbeck2001}.
It is known that calculated $\ku$'s of $\yc$-based alloy may be systematically more diminutive than their experimental counterparts.

The number of $k$-points in the first Brillouin zone is 2197.
The maximum orbital angular momentum cutoff $l_{\rm max}$ of each element was set to 3.
The width of the energy contour for complex integration was set to 1.0 Ry, where $4p$ orbitals of yttrium are treated as the core.

The space group of hexagonal CaCu$_5$-type YCo$_5$ is No.191 ($P6/mmm$), and cobalt atoms occupy two inequivalent Wyckoff positions $3g$- and $2c$-sites.
From a crystallographic point of view, the spatial volume occupied by the $2c$-site is smaller than that of the $3g$-site.
The atomic radius of nickel (0.124 \AA) is smaller than that of cobalt (0.125 \AA), and thus nickel can easily enter the $2c$-site \cite{Chuang1982,Deportes1976,Yamada1999}. 
In contrast, the atomic radii of iron and copper are 0.126 \AA\ and 0.128 \AA, respectively.
Hence, iron and copper can occupy the $3g$-site \cite{Chuang1982}.

We considered 726 different disordered compounds; $\ycofecuni$, where the concentration $x,y$, and $z$ are varied by 0.1 under $0\leq x+y\leq 1$ and $0\leq x,y,z \leq 1$.
We optimized the lattice parameters of YCo$_5$, YCo$_3$Ni$_2$, YFe$_3$Co$_2$, YFe$_3$Ni$_2$, YCu$_3$Co$_2$, and YCu$_3$Ni$_2$ within LDA using the Vienna Ab initio Simulation Package (VASP) \cite{Kresse1996}, and summarized the lattice parameters in \retable{tb:lat}.
The relative change of the lattice parameter with the addition of the third element is consistent with the relationship between the large and small atomic radii mentioned above.
Lattice parameters for the remaining compounds were determined using Vegard's law with optimized values of the above six materials.
The determined lattice parameters are reasonable except for Y(Co$_{1-z}$Ni$_z$)$_5$ ($0.4<z$), where the measured lattice parameters are not monotonically changed with increasing nickel content \cite{Chuang1982,Ishikawa2003}.
Vegard's law is also applicable to the lattice constant of iron doping cases according to the experimental values of YCo$_{5-x}$Fe$_x$ ($x<0.5$) \cite{Maruyama1999}.
It is revealed that the lattice constant determined using LDA is typically underestimated owing to the overestimation of the binding energy of $3d$ orbitals.
In addition to LDA calculations, we also performed GGA calculations to confirm the framework's accuracy.

The magnetocrystalline anisotropy energy is defined as $\ku = E_{100}-E_{001}$, where $E_{100}$ ($E_{001}$) is the total energy of alloys magnetized along the $[100]$ ($[001]$) direction in a hexagonal unit cell. The positive $\ku$ corresponds to the uniaxial anisotropy.
The dependence of the calculated $\ku$ on the lattice parameters is discussed in \resec{sec:yco5}.
The $3g$-site in the $P6/mmm$ is separated into three inequivalent sites when the quantization axis is rotated by 90 degrees.

\begin{center}
\begin{table*}[t!]
\caption{Lattice parameters optimized by VASP with the LDA functional. The uniaxial magnetocrystalline anisotropy energy ($\ku$), saturation magnetization ($\js$), and anisotropy fields ($\ha$) of $\yc$ and pseudo-binary alloys calculated using the full-potential KKR method.
Except for YCu$_3$Ni$_2$, all listed compounds are ferromagnetic.}

\label{tb:lat}
\scalebox{0.85}{
\begin{tabular}{lcccccccccc}
\hline\hline
              &    $a$ [\AA]&     $c$ [\AA]&        $c/a$& Volume [\AA$^3$]&   $\ku$ [meV/f.u.]& $\ku$ [MJ/m$^3$]&       $\js$ [T]&    $\ha$ [T] \\
\hline
YCo$_5$       &       4.749&        3.787&        0.798&        73.98&         0.84&         1.82&         1.03&         4.49  \\ 
YFe$_3$Co$_2$ &       4.756&        3.872&        0.814&        75.82&         2.37&         5.00&         1.12&         11.42 \\
YCu$_3$Co$_2$ &       4.785&        3.980&        0.832&        78.91&         1.72&         3.48&         0.27&         33.34 \\ 
YCo$_3$Ni$_2$ &       4.724&        3.834&        0.812&        74.11&         0.49&         1.05&         0.65&         4.11  \\ 
YFe$_3$Ni$_2$ &       4.730&        3.919&        0.828&        75.93&         0.29&         0.62&         0.78&         2.01  \\ 
YCu$_3$Ni$_2$ &       4.760&        4.027&        0.846&        79.01&          ---&          ---&         0.00&         ---   \\ 
\hline\hline
\end{tabular}
}
\end{table*}
\end{center}

\begin{figure}[b!]
\centering
\includegraphics[width=\linewidth]{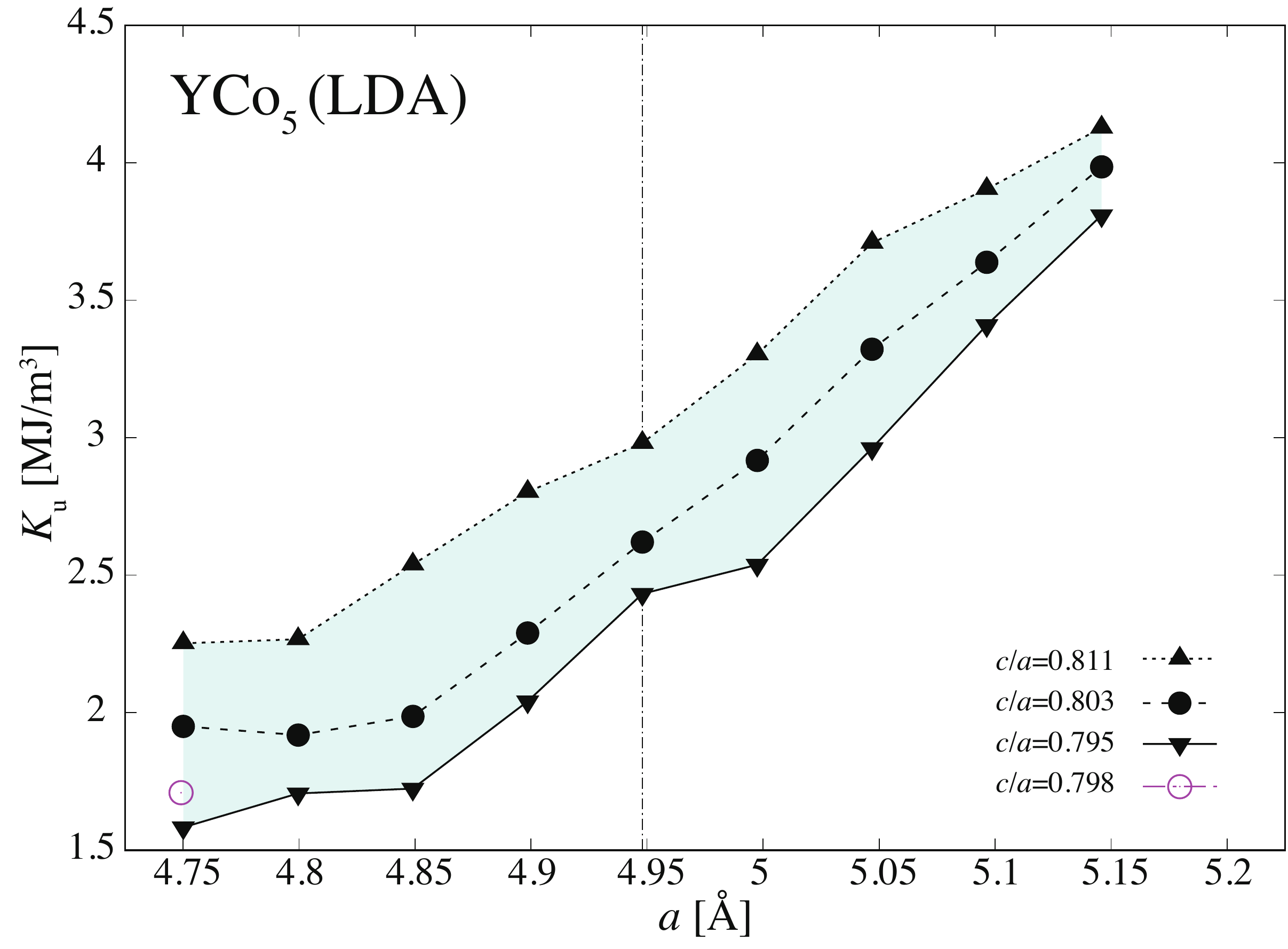}
\caption{Calculated $\ku$ of $\yc$ by full-potential KKR with the LDA functional as the function of the lattice parameter $a$.  
The filled circle is the $\ku$ calculated using experimental $c/a=0.8034$ \cite{Schweizer1980}, and the triangles are the $\ku$ calculated using changed $c/a$ by $\pm$1\% from the experimental value.
The vertical dotted line is the experimental lattice constant, which was found to be $a=4.948$ \AA\ \cite{Schweizer1980}.
The open circle is the $\ku$ calculated using the optimized lattice parameters shown in \retable{tb:lat}. } 
\label{fig:mae_yco5}
\end{figure}

\begin{figure*}[t!]
\centering
\includegraphics[width=\linewidth]{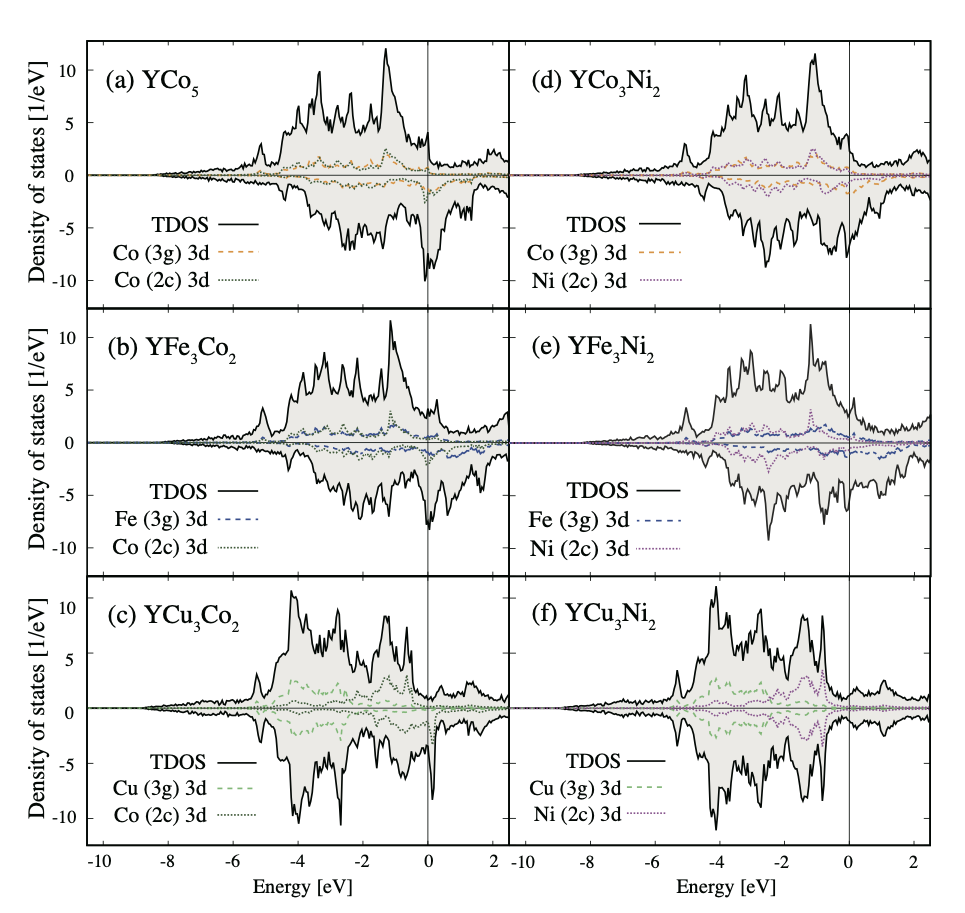}
\caption{Calculated total and projected density of states of the transition metal $3d$ orbital for (a) $\yco$, (b) $\yfeco$, (c) $\ycuco$, (d) $\ynico$, (e) $\yfeni$, and (f) $\ycuni$ using the full-potential KKR method with the LDA functional.
Fermi level is set to zero. It is shown that $\ycuni$ is in paramagnetic states with $\js=0$ T.}
\label{fig:dos}
\end{figure*}

\section{Results and Discussions}
\subsection{$\yc$}\label{sec:yco5}
In \retable{tb:lat}, the optimized lattice volume of $\yc$ in LDA was 73.9 \AA, which is close to the volume of the first-order Lifshitz transition \cite{Rosner2006}, but the calculated electronic structure was in a high-spin state with a total magnetic moment of 6.52 $\mb$/f.u. ($\js=1.03$ T).
The moment was smaller than the experimental values, which were in the range of 7.9-8.3 $\mb$/f.u. \cite{Tatsumoto1971,Frederick1974,Klein1974}. This might be probably due to the less orbital polarization in the calculation.

As mentioned in Refs.\cite{Asali2019,Ucar2020}, anisotropy is strongly affected by both the crystal structures and choice of exchange-correlation functional.
\refig{fig:mae_yco5} shows the lattice parameter dependence of $\ku$ using FPKKR with the LDA functional, where both parameters $a$ and $c/a$ are changed. $\ku$ increases as $a$ increases.
The $\ku$ calculated with optimized lattice parameters ($a=4.749$ \AA, $c/a=0.798$) was smaller than the $\ku$ calculated with the experimental parameters ($a=4.948$ \AA, $c/a=0.803$ \cite{Schweizer1980}).
Even though we used the experimental parameters, the calculated $\ku$ was underestimated compared to the experimental $\ku$.
This underestimation is related to the underestimated orbital polarization of cobalt \cite{Schweizer1980,Heidemann1975}.
In the subsequent calculations, we used the optimized parameters and their interpolated values by using Vegard's law to perform the calculations in a consistent way.

\refig{fig:dos}(a) shows the calculated total density of states of $\yc$, along with the projected density of states for $3d$ components of cobalt atoms.
In $\yc$, strong ferromagnetism was realized, as in hcp cobalt.
The shape of the density of states was similar to those reported in previous studies \cite{Rosner2006,Plugaru2014,Burzo2020}.
In the minority spin states, the peak at the Fermi level mainly contributes to cobalt $3d$ states.
In \refig{fig:dos} (b)-(f), the calculated total density of states for doped $\yco$ are shown together with the projected density of states of the $3d$ orbital of transition metals at both $2c$- and $3g$-sites.
Apart from $\ycuni$, all the compounds are in the ferromagnetic configuration with the finite magnetization listed in \retable{tb:lat}.
\refig{fig:dos} (c) and (d) show strong ferromagnetic character in spite of the addition of third elements in contrast to the cases of \refig{fig:dos} (e) and (f).
In $\yfeni$ of \refig{fig:dos}(e), all cobalt has been substituted by other elements, and no distinct large exchange splitting can be identified.
In \refig{fig:dos}(f), the Fermi level rises and almost all $3d$-bands are occupied ($3d^{8.76}$ and $3d^{9.48}$ for nickel and copper), leading to a non-magnetic state in $\ycuni$.

\begin{figure*}[t!]
\centering
\includegraphics[width=\linewidth]{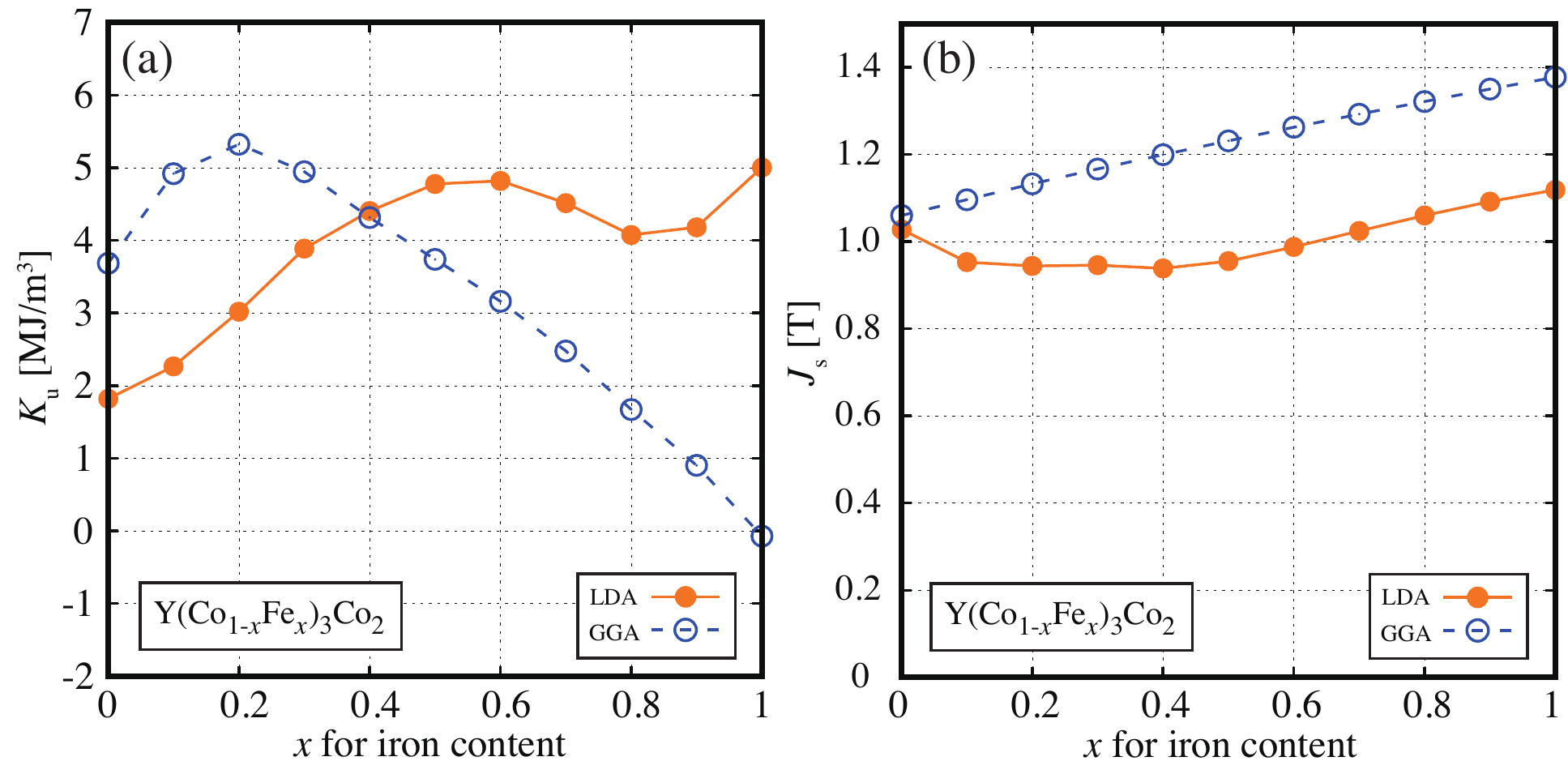}
\caption{(a) Calculated uniaxial anistoropy constant $\ku$ and (b) magnetization $\js$ of Y(Co$_{1-x}$Fe$_x$)$_3$Co$_2$ as a function of iron content using the full-potential KKR method with LDA and GGA.
Calculated values of $\yco$ are $\ku=1.82$ (3.68) $\mjmt$ and $\js=1.03$ (1.06) T for LDA (GGA).
}
\label{fig:mae_js_fe}
\end{figure*}

\subsection{$\ycofe$}
\refig{fig:mae_js_fe}(a) shows the calculated $\ku$ of Y(Co$_{1-x}$Fe$_x$)$_3$Co$_2$ as a function of iron content using FPKKR with LDA and GGA.
The LDA result shows two peaks of $\ku$ at $x\sim0.5$ and $x\sim1.0$.
The increase in the anisotropy in low concentrations of iron ($x<0.5$) is plausibly illustrated in the crystal field theory \cite{Inomata1981}.
According to this model, cobalt atoms at the $3g$-site contribute to the planer anisotropy and replacing them with iron atoms causes a large uniaxial anisotropy.
This has been confirmed for iron at the 3g-site is selectively replaced with cobalt, as well as with uniform replacement at both the $2c$- and $3g$-sites \cite{Patrick2019}.
The first peak is also found for rigid band calculations \cite{Steinbeck2001}, and the calculations employing virtual crystal approximation \cite{Larson2003}.

However, the experimental peak is found at around $x=0.17$ \cite{Franse1988}, which is smaller than the present calculation.
The possible reasons for this discrepancy are threefold: (1) the site preference of iron atoms, (2) the different lattice constants used in the present calculation, and (3) the choice of exchange-correlation functional.
Firstly, the iron-site dependence of $\ku$ in YCo$_{5-x}$Fe$_x$ ($x<0.1$) is theoretically investigated in Ref. \cite{Patrick2019}, where the $\ku$ with iron at the $3g$-site is more significant than that at the $2c$-site.
The second is that the dependency of $\ku$ on the iron concentration is different from that with larger lattice constants.
The LDA calculation with GGA optimized structures has a peak at $x=0.2$ with the FPKKR method (not shown).
A peak at $x\sim 0.2$ is reported in the LDA calculation by using the lattice constant obtained from experiments \cite{Patrick2019}.
The third reason is evident from \refig{fig:mae_js_fe}(a).
In the GGA calculation, the peak found at $x=0.2$ is moved to a lower concentration region than the peak position of LDA. Hence, the peak position is located on the high iron concentration side only when the LDA calculation is performed using the lattice constants optimized with LDA.

In the GGA case of $\yc$, the absolute value of $\ku=3.68$ $\mjm$ ($a=4.91$ \AA, $c/a=0.80$) is higher than that of the LDA even when the expanded lattice constants are used, as shown in \refig{fig:mae_yco5}.
In \refig{fig:mae_js_fe}(a), it is shown that there is only one peak in the GGA calculation.
In the region above $x=0.2$, the anisotropy constant monotonically decreases with increasing iron concentration.
From the GGA calculations, the uniaxial anisotropy of YFe$_3$Co$_2$ is weaker than that of YCo$_5$.
The lower value of anisotropy in YFe$_3$Co$_2$ is also confirmed by earlier theoretical calculations using GGA \cite{Landa2020}.

\refig{fig:mae_js_fe}(b) shows the calculated $\js$ as a function of iron content.
Unexpectedly, in the LDA, $\js$ decreases with increasing iron until $x=0.4$; the reason for this is not identified.
However, this trend is not found in the GGA calculation, where the $\js$ monotonically increases and is more significant than $\js$ in the LDA case.
Therefore, the decrease in magnetization in the low iron concentration is only seen for the LDA calculations and small lattice constants.
In the case of LDA, higher iron concentrations ($x>0.4$) increase the magnetization, which can be attributed to iron's more considerable spin magnetic moments, like Fe-Co binary alloys.
In Y(Co$_{1-x}$Fe$_x$)$_5$ system, we confirmed that the peak of the Slater-Pauling curve appears when the number of $d$-electrons larger than that of YFe$_3$Co$_2$.
In addition, the peak has been theoretically predicted in YCo$_4$Fe \cite{Larson2004}.

\begin{figure*}[t!]
\centering
\includegraphics[width=\linewidth]{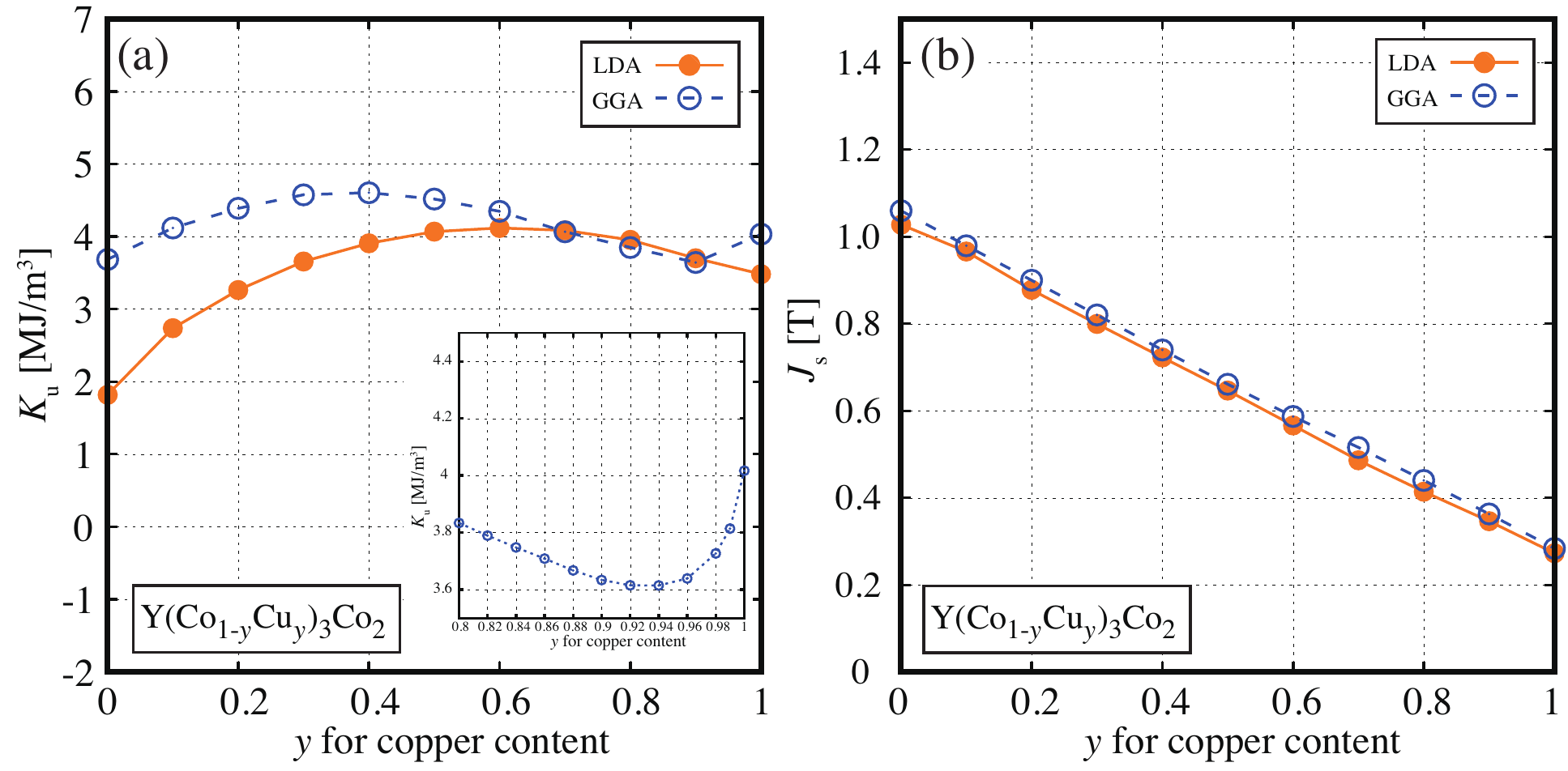}
\caption{(a) Calculated uniaxial anisotropy constant $\ku$ and (b) magnetization $\js$ of Y(Co$_{1-y}$Cu$_y$)$_3$Co$_2$ as a function of copper content using the full-potential KKR method with LDA and GGA.
The inset in (a) shows the detailed copper dependence of $\ku$ at high copper concentration using GGA.
}
\label{fig:mae_js_cu}
\end{figure*}

\subsection{$\ycocu$}\label{subsec:ycc}
In \retable{tb:lat}, the uniaxial anisotropy remains in the YCu$_3$Co$_2$ ($y=1.0$), where cobalt atoms at the $3g$-site are fully substituted for copper.
From a crystallographic point of view, there are two layers: one includes the $3g$-site with copper, and the other includes yttrium and the $2c$-site with cobalt.
Since the local moment of copper is negligibly small (0.001 $\mb$), the ferromagnetism in the system is carried by the cobalt with the moment of 0.94 $\mb$.
Therefore, the anisotropy of the material is mainly due to the magnetic interaction within the layer containing the $2c$-site, not by the magnetic interaction along the perpendicular direction.

\refig{fig:mae_js_cu}(a) shows the calculated $\ku$ of Y(Co$_{1-y}$Cu$_y$)$_3$Co$_2$ as a function of copper by using LDA and GGA. In the LDA, $\ku$ of $\ycocu$ increases as copper increases until it takes maximum at $y\sim0.6$.
This contradicts the fact that the experimental $\ku$ of YCo$_3$Cu$_2$ ($y=0.67$) is smaller than the $\ku$ of $\yco$ \cite{Tellez2000}.
This discrepancy may be due to the preferential site occupancy of copper.
In the low copper concentration, however, the theoretical calculations revealed that a larger $\ku$ is caused by adding copper to the $2c$-site rather than the $3g$-site \cite{Patrick2019}.
Hence, the small experimental value is probably not explained by the site preference of copper alone.

Compared to the LDA calculation, the peak in the GGA calculation is shifted to the lower copper concentration, like the iron-doped case in \refig{fig:mae_js_fe}(a).
At $y\sim0.67$, the calculated $\ku$ is still larger than that of pristine $\yc$ ($y=0$).
As already mentioned, this enhancement of anisotropy contradicts the experimental results in the LDA cases \cite{Tellez2000}.
In contrast, we can estimate $\ku$ of YCo$_4$Cu from the anisotropy field ($\ha$) at 0 K, which is extrapolated from the $\ha$ from the finite temperature \cite{Tellez2000,Patrick2019}.
The experiment indicates that the $\ku$ is enhanced by copper addition, at least in the low copper concentration region ($y\leq0.33$).
Even though the peak locations are different, the present LDA and GGA calculations also confirm the increase in the anisotropy. The inset figure in \refig{fig:mae_js_cu}(a) shows the detailed $\ku$ at high copper concentrations from GGA.
A minimum is observed at $y\sim 0.93$, and the further increase in copper enhances the $\ku$.
The calculated $\ku=3.6$ $\mjmt$ at $y=0.93$ is a minimum value but is still comparable to the $\ku$ of $\yc$.
However, such material with high copper concentration is unsuitable for magnets because the magnetization becomes extremely small.
The magnetization calculated for LDA and GGA, shown in \refig{fig:mae_js_cu}(b), decreases drastically as the copper content increases.

\begin{figure*}[t!]
\centering
\includegraphics[width=\linewidth]{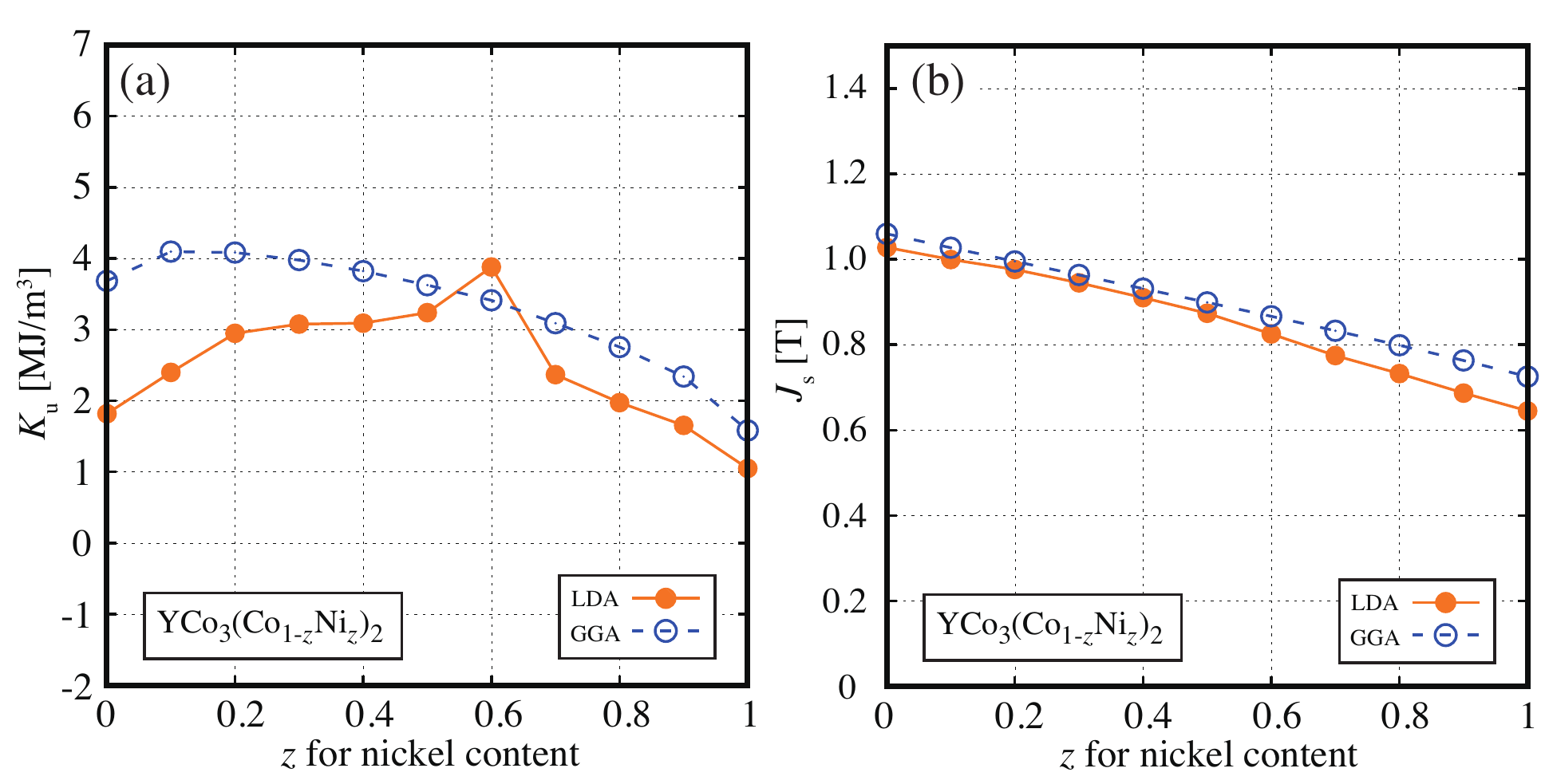}
\caption{(a) Calculated uniaxial anistoropy constant $\ku$ and (b) magnetization $\js$ of YCo$_3$(Co$_{1-z}$Ni$_z$)$_2$ as a function of nickel content obtained using the full-potential KKR method with LDA and GGA.
}
\label{fig:mae_js_ni}
\end{figure*}

\subsection{$\yconi$}\label{subsec:ycn}
\refig{fig:mae_js_ni}(a) shows the $\ku$ of YCo$_3$(Co$_{1-z}$Ni$_z$)$_2$ as a function of nickel calculated using LDA and GGA.
The chemical trend in LDA is almost the same as that of GGA except for the anomalous value at $z=0.6$.
The calculations show that a small amount of nickel enhances the $\ku$.
A similar enhancement in low nickel concentration can be theoretically obtained in Gd(Co$_{1-z}$Ni$_z$)$_5$ \cite{Tedstone2019}.
The theoretical calculation indicates that the nickel at the $2c$-site enhances the $\ku$ in the gadolinium system.
For yttrium, the enhancement of anisotropy has been analyzed using the band filling model in the rigid band picture \cite{Daalderop1996}.
An enhancement in anisotropy by the electron addition is similar to the case of copper shown in \refig{fig:mae_js_cu}(a).
However, the experiment \cite{Deportes1976,Buschow1976} demonstrated the monotonic decrease of $\ku$ as nickel content increases.
This discrepancy is attributed to the preferential occupation of nickel.
As the nickel concentration increases, it occupies the $3g$-site as well.
In fact, the neutron experiment indicates that the composition of $x=0.2$ for Y(Co$_{1-x}$Ni$_x$)$_5$ is Y(Co$_{0.86}$Ni$_{0.14}$)$_3$(Co$_{0.71}$Ni$_{0.29}$)$_2$ \cite{Deportes1976}.
Therefore, the present assumption that the nickel occupied only the $2c$-site does not hold for higher nickel concentration.

\refig{fig:mae_js_ni}(b) shows $\js$ of $\yconi$ calculated by using LDA and GGA.
The values of $\js$ monotonically decrease as the nickel content increases. This dependency corresponds to the slope of the right half of the Slater-Pauling curve, as in the cases of copper addition.
The present calculation underestimates the magnetization as compared to the experiment \cite{Ishikawa2003} but qualitatively reproduces the nickel concentration dependence of the experimental magnetization.

In contrast to iron addition in \refig{fig:mae_js_fe}(b), the calculated magnetization shows less significant differences between GGA and LDA.
This might be due to the existence of the significant states near the Fermi level in the density of states of $\yc$ shown in \refig{fig:dos}(a). Iron substitution shifts the Fermi energy to lower energy, which significantly changes the contribution of the significant part and affects the anisotropy in the case of LDA (\refig{fig:dos}(b)).
In contrast, nickel substitution can suppress this effect, which increases the number of $d$-electrons and compensates for the decrease in the number of $d$-electrons induced by iron doping.
Thus, if there is some nickel ($e.g.$ $z=0.2$ for Y(Co$_{1-x}$Fe$_x$)$_3$(Co$_{1-z}$Ni$_z$)$_2$), it is expected that the dependency of $\ku$ on iron concentration would be moderate (shown in Fig. S1(a) in supplemental materials \cite{suppl}).
In GGA, the dominant states are in a deeper energy region (density of states with GGA in Fig. S7 in supplemental materials \cite{suppl}).
In the case of iron, therefore, the calculated $\js$ of GGA are significantly different from LDA cases.

It is worth mentioning the anomalous value of the LDA calculation at $z=0.6$.
In contrast to the calculation with higher nickel content \cite{Yamada1999}, the spin-state transition (the first-order Lifshitz transition) does not occur at $z=0.6$.
In the present LDA calculation,
\refig{fig:dos}(a) and (d) indicate that both $\yc$ and YCo$_3$Ni$_2$ are in the high-spin states.
In the intermediate concentration range, we also confirmed that YCo$_3$(Co$_{1-z}$Ni$_z$)$_2$ are still in the high-spin states (local density of states of $3d$ orbitals in Fig. S3 in supplemental materials \cite{suppl}).
At $z=0.6$, the volume dependence of the magnetic moment also indicates that the system is in the high-spin state at the volume determined by Vegard's law (Fig. S4).
In spite of no transitions, \refig{fig:mae_js_ni}(b) shows that the differential coefficient of magnetization with nickel concentration varies at $z=0.6$.
The changes in the differential coefficients might be related to anomalies in the $\ku$.
It is necessary to calculate the contribution to the magnetocrystalline anisotropy from the band structure to investigate this, but this is beyond the scope of this paper.

\begin{figure}[t!]
\centering
\includegraphics[width=\linewidth]{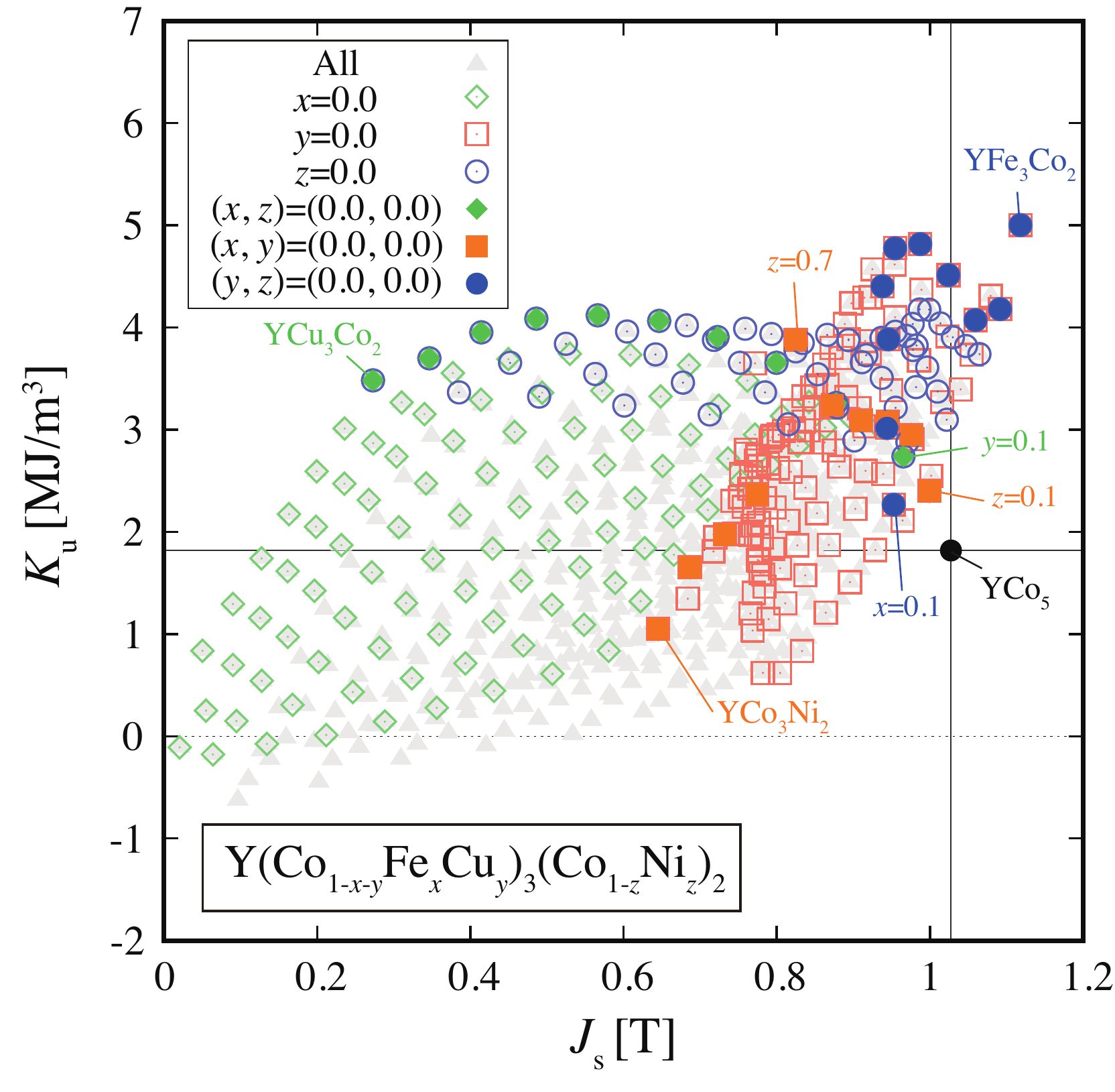}
\caption{Calculated magnetization $\js$ and anisotropy $\ku$ constant of alloys excluding a few paramagnetic alloys, such as YCu$_3$Ni$_2$, by the full-potential KKR method with LDA functional.
The triangles correspond to all materials, the open diamonds to $x=0.0$, the open squares to $y=0.0$, the open circles to $z=0.0$, the filled diamonds to $(x,z)=(0.0,0.0)$, the filled squares to $(x,y)=(0.0,0.0)$, and the filled circles to $(y,z)=(0.0,0.0)$.
The vertical and horizontal lines represent the $\ku=1.82$ $\mjmt$ and $\js=$ 1.03 T of $\yc$, respectively.}
\label{fig:mae_js}
\end{figure}

\begin{figure*}[t!]
\centering
\includegraphics[width=\linewidth]{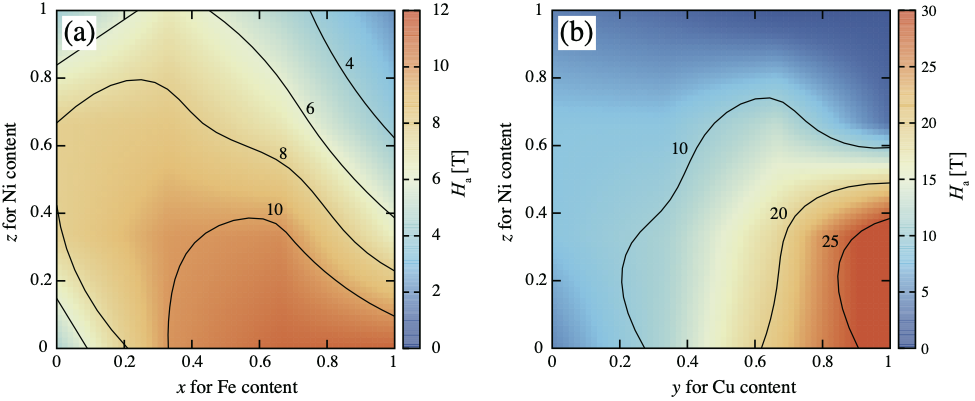}
\caption{Calculated $\ha=2\ku/\js$ as functions of impurity contents by the full-potential KKR method with the LDA functional for (a) $\ycofeni$ and (b) $\ycocuni$.
The LDA calculation shows the $\ha=4.49$ T of $\yc$ (bottom left).
}
\label{fig:ha}
\end{figure*}

\subsection{Magnetization vs anisotropy constant}
\refig{fig:mae_js} shows the calculated $\js$ and $\ku$ of $\ycofecuni$ by LDA except for paramagnetic materials, such as YCu$_3$Ni$_2$ (the contour plots can be available in supplemental materials \cite{suppl}). 
The intersection of the solid lines represents the values of $\yc$.
The magnetization decreases by adding a small amount of the third element, but the anisotropy increases.
\refig{fig:ha}(a) shows the calculated $\ha=2\ku/\js$ of doped $\yc$ without copper as a function of iron and nickel content.
The $\ha$ of the disordered alloy is more significant than $\ha=4.49$ T of $\yc$ in a relatively wide range.

\refig{fig:ha}(b) shows the calculated $\ha$ of doped $\yc$ without iron as a function of copper and nickel content.
The copper doping also enhances $\ha$ for the low nickel concentration region because the $\js$ drastically decreases when added copper.
This agrees with the previous calculation assuming no site-preference of copper \cite{Patrick2019}.
As mentioned in \resec{subsec:ycc}, $\ha$ in higher copper concentration is different from the experimental result \cite{Tellez2000}.
In addition to the site preference of copper, another origin of the discrepancy is a way to calculate $\ha=2\ku/\js$ in the calculation.
In the region of large copper concentration, where the magnetic moment is considerably low, it is not appropriate to determine the anisotropic magnetic field using the formula $\ha=2\ku/\js$.
Consequently, the large $\ha=33.34$ T of YCu$_3$Co$_2$ should not be observed.

In \refig{fig:mae_js}, when nickel or copper alone is added, $\js$ and $\ku$ simultaneously increase.
The most promising candidate is $\yfeco$, located at the upper right of $\yc$ in the figure; herein, both $\ku$ and $\js$ are more significant than those of $\yco$.
The 1-5 phases of YFe$_3$Co$_2$ are unstable because iron is unlikely to be dissolved in the 1-5 phases as a solid solution.
However, in the previous theoretical calculations for the samarium case, it was shown that nickel stabilized the 1-5 phase and iron-rich SmCoFeNi$_3$ was proposed \cite{Soderlind2017}.

\begin{figure}[t!]
\centering
\includegraphics[width=\linewidth]{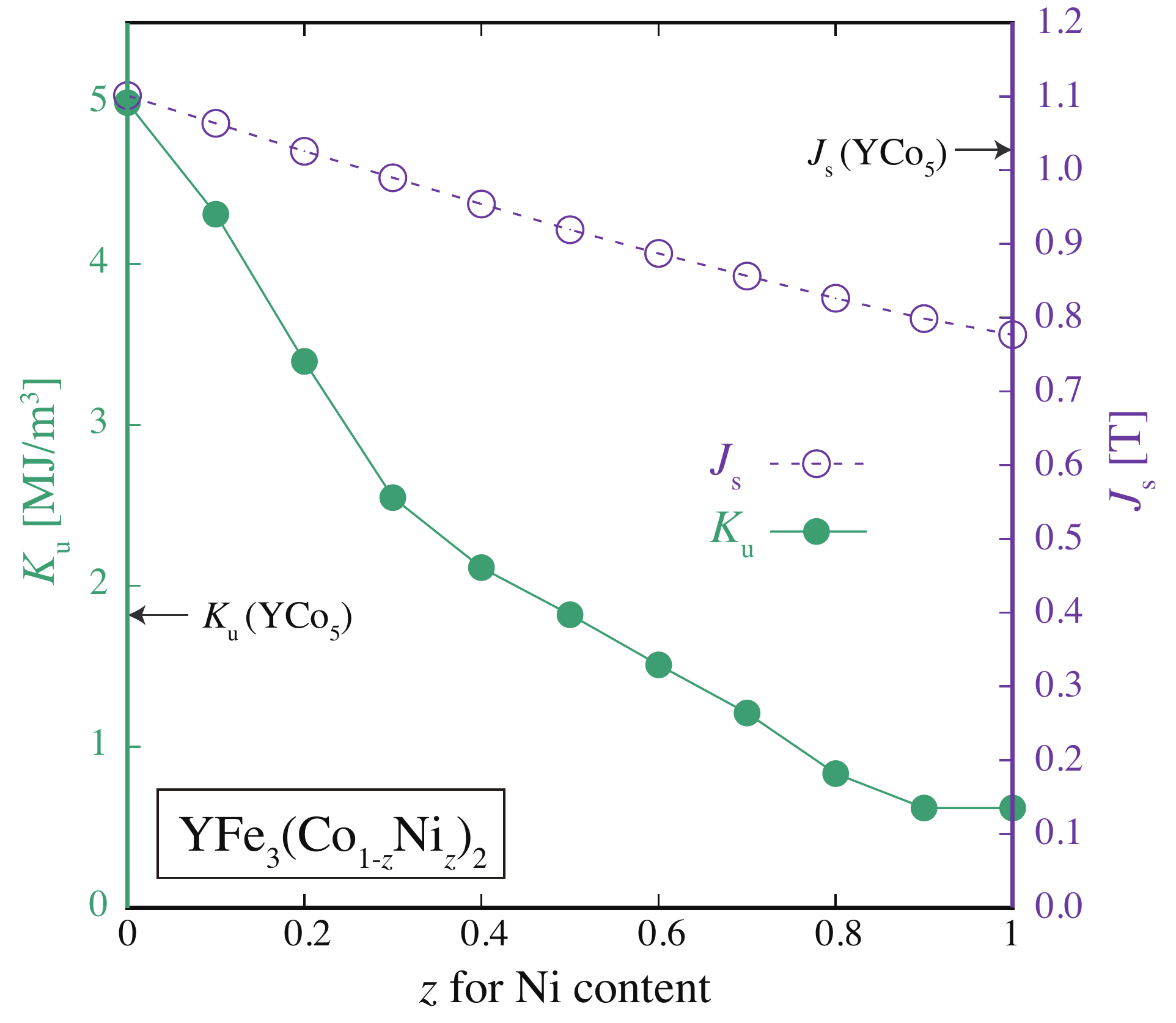}
\caption{Calculated $\ku$ and $\js$ of YFe$_3$(Co$_{1-z}$Ni$_z$)$_2$ as a function of nickel content using the full-potential KKR method with the LDA functional.
Filled circle is $\ku$ and open circle is $\js$. Arrows are values of $\yc$ as a reference.}
\label{fig:mae_js_yfcn}
\end{figure}

Additionally, Landa $et$ $al.$ concluded that nickel stabilized YFe$_3$(Co$_{1-z}$Ni$_z$)$_2$ from the calculation of phase diagrams (CALPHAD) \cite{Landa2020}.
According to this result, higher nickel content enhances the stabilization of the 1-5 phases.
We calculated $\ku$ and $\js$ using the FPKKR to investigate the magnetic properties of YFe$_3$(Co$_{1-z}$Ni$_z$)$_2$. The results are shown in \refig{fig:mae_js_yfcn}. The calculated $\ku$ decreases as nickel content increases, which is in line with the earlier theoretical calculations by Landa $et$ $al.$.

\refig{fig:mae_js_yfcn} indicates that nickel, as the stabilization element, will deteriorate both the $\ku$ and $\js$.
Therefore, the magnetic performance and stabilization for the 1-5 phases are in a trade-off relation, and optimal nickel content should be determined for practical use.
The arrows represent the reference values for $\yc$ in the figure, and the magnetization is greater than that of $\yc$ even when nickel is added up to $z=0.2$.
The anisotropy is also suggested to have a larger $\ku$ than that of $\yc$ up to $z=0.5$.
From \refig{fig:ha}(a), we can expect the $\ha$ of YFe$_3$(Co$_{1-x}$Ni$_x$)$_2$ to be equal to or higher than that of $\yc$.
As originally mentioned in Ref. \cite{Landa2020}, if the 1-5 phase can exist stably, YFe$_3$(Co$_{1-x}$Ni$_x$)$_2$ can be a useful magnet with better performance than $\yc$.

\section{Summary}
We calculated the magnetocrystalline anisotropy and magnetization of Y(Co,Fe,Cu,Ni)$_5$ in order to determine its optimal composition as a permanent magnet. The systematic calculations are based on the full-potential KKR Green's function method combined with the coherent potential approximation.
The calculated anisotropy of $\yc$ strongly depended on lattice parameters. Optimized lattice parameters were used for consistency.

The results obtained using LDA indicate that the anisotropy constant $\ku$ and magnetization are higher than $\yc$ when the iron is added, and they reach a maximum in YFe$_3$Co$_2$.
However, nickel, used to stabilize the 1-5 phases, decreases the magnetization and anisotropy constant.
The magnetic anisotropy and magnetization exhibit higher values than $\yc$ up to $z=0.4$ and $z=0.2$.
The calculated anisotropy field is expected to be larger than the original $\yc$ up to $z=0.5$.

The present systematic calculations shed light on the magnetic properties and anisotropy of $\yc$ systems. The underestimation of the magnetocrystalline anisotropy constant is due to the absence of orbital polarization enhancement.
The anisotropy constant obtained by the systematic calculations in the $\yc$ system is also essential for discussing the contribution of the cobalt sublattice to the anisotropy of the SmCo$_5$ system.

\bibliography{citation}

\begin{thebibliography}{51}%
\makeatletter
\providecommand \@ifxundefined [1]{%
 \@ifx{#1\undefined}
}%
\providecommand \@ifnum [1]{%
 \ifnum #1\expandafter \@firstoftwo
 \else \expandafter \@secondoftwo
 \fi
}%
\providecommand \@ifx [1]{%
 \ifx #1\expandafter \@firstoftwo
 \else \expandafter \@secondoftwo
 \fi
}%
\providecommand \natexlab [1]{#1}%
\providecommand \enquote  [1]{``#1''}%
\providecommand \bibnamefont  [1]{#1}%
\providecommand \bibfnamefont [1]{#1}%
\providecommand \citenamefont [1]{#1}%
\providecommand \href@noop [0]{\@secondoftwo}%
\providecommand \href [0]{\begingroup \@sanitize@url \@href}%
\providecommand \@href[1]{\@@startlink{#1}\@@href}%
\providecommand \@@href[1]{\endgroup#1\@@endlink}%
\providecommand \@sanitize@url [0]{\catcode `\\12\catcode `\$12\catcode
  `\&12\catcode `\#12\catcode `\^12\catcode `\_12\catcode `\%12\relax}%
\providecommand \@@startlink[1]{}%
\providecommand \@@endlink[0]{}%
\providecommand \url  [0]{\begingroup\@sanitize@url \@url }%
\providecommand \@url [1]{\endgroup\@href {#1}{\urlprefix }}%
\providecommand \urlprefix  [0]{URL }%
\providecommand \Eprint [0]{\href }%
\providecommand \doibase [0]{http://dx.doi.org/}%
\providecommand \selectlanguage [0]{\@gobble}%
\providecommand \bibinfo  [0]{\@secondoftwo}%
\providecommand \bibfield  [0]{\@secondoftwo}%
\providecommand \translation [1]{[#1]}%
\providecommand \BibitemOpen [0]{}%
\providecommand \bibitemStop [0]{}%
\providecommand \bibitemNoStop [0]{.\EOS\space}%
\providecommand \EOS [0]{\spacefactor3000\relax}%
\providecommand \BibitemShut  [1]{\csname bibitem#1\endcsname}%
\let\auto@bib@innerbib\@empty
\bibitem [{\citenamefont {Sagawa}\ \emph {et~al.}(1984)\citenamefont {Sagawa},
  \citenamefont {Fujimura}, \citenamefont {Togawa}, \citenamefont {Yamamoto},\
  and\ \citenamefont {Matsuura}}]{Sagawa1984}%
  \BibitemOpen
  \bibfield  {author} {\bibinfo {author} {\bibfnamefont {M.}~\bibnamefont
  {Sagawa}}, \bibinfo {author} {\bibfnamefont {S.}~\bibnamefont {Fujimura}},
  \bibinfo {author} {\bibfnamefont {N.}~\bibnamefont {Togawa}}, \bibinfo
  {author} {\bibfnamefont {H.}~\bibnamefont {Yamamoto}}, \ and\ \bibinfo
  {author} {\bibfnamefont {Y.}~\bibnamefont {Matsuura}},\ }\href
  {https://doi.org/10.1063/1.333572} {\bibfield  {journal} {\bibinfo  {journal}
  {J. Appl. Phys.}\ }\textbf {\bibinfo {volume} {55}},\ \bibinfo {pages} {2083}
  (\bibinfo {year} {1984})}\BibitemShut {NoStop}%
\bibitem [{\citenamefont {Croat}\ \emph {et~al.}(1984)\citenamefont {Croat},
  \citenamefont {Herbst}, \citenamefont {Lee},\ and\ \citenamefont
  {Pinkerton}}]{Croat1984}%
  \BibitemOpen
  \bibfield  {author} {\bibinfo {author} {\bibfnamefont {J.~J.}\ \bibnamefont
  {Croat}}, \bibinfo {author} {\bibfnamefont {J.~F.}\ \bibnamefont {Herbst}},
  \bibinfo {author} {\bibfnamefont {R.~W.}\ \bibnamefont {Lee}}, \ and\
  \bibinfo {author} {\bibfnamefont {F.~E.}\ \bibnamefont {Pinkerton}},\ }\href
  {\doibase 10.1063/1.333571} {\bibfield  {journal} {\bibinfo  {journal} {J.
  Appl. Phys.}\ }\textbf {\bibinfo {volume} {55}},\ \bibinfo {pages} {2078}
  (\bibinfo {year} {1984})}\BibitemShut {NoStop}%
\bibitem [{\citenamefont {Coey}(2020)}]{Coey2020}%
  \BibitemOpen
  \bibfield  {author} {\bibinfo {author} {\bibfnamefont {J.~M.~D.}\
  \bibnamefont {Coey}},\ }\href {\doibase 10.1016/j.eng.2018.11.034} {\bibfield
   {journal} {\bibinfo  {journal} {Eng.}\ }\textbf {\bibinfo {volume} {6}},\
  \bibinfo {pages} {119} (\bibinfo {year} {2020})}\BibitemShut {NoStop}%
\bibitem [{\citenamefont {Tatsumoto}\ \emph {et~al.}(1971)\citenamefont
  {Tatsumoto}, \citenamefont {Okamoto}, \citenamefont {Fujii},\ and\
  \citenamefont {Inoue}}]{Tatsumoto1971}%
  \BibitemOpen
  \bibfield  {author} {\bibinfo {author} {\bibfnamefont {E.}~\bibnamefont
  {Tatsumoto}}, \bibinfo {author} {\bibfnamefont {T.}~\bibnamefont {Okamoto}},
  \bibinfo {author} {\bibfnamefont {H.}~\bibnamefont {Fujii}}, \ and\ \bibinfo
  {author} {\bibfnamefont {C.}~\bibnamefont {Inoue}},\ }\href {\doibase
  https://doi.org/10.1051/jphyscol:19711186} {\bibfield  {journal} {\bibinfo
  {journal} {J. Phys. Colloq.}\ }\textbf {\bibinfo {volume} {32}},\ \bibinfo
  {pages} {C1} (\bibinfo {year} {1971})}\BibitemShut {NoStop}%
\bibitem [{\citenamefont {Coey}(2011)}]{Coey2011}%
  \BibitemOpen
  \bibfield  {author} {\bibinfo {author} {\bibfnamefont {J.~M.~D.}\
  \bibnamefont {Coey}},\ }\href {\doibase 10.1109/TMAG.2011.2166975} {\bibfield
   {journal} {\bibinfo  {journal} {IEEE Trans. Magn.}\ }\textbf {\bibinfo
  {volume} {47}},\ \bibinfo {pages} {4671} (\bibinfo {year}
  {2011})}\BibitemShut {NoStop}%
\bibitem [{\citenamefont {Alameda}\ \emph {et~al.}(1981)\citenamefont
  {Alameda}, \citenamefont {Givord}, \citenamefont {Lemaire},\ and\
  \citenamefont {Lu}}]{Alameda1981}%
  \BibitemOpen
  \bibfield  {author} {\bibinfo {author} {\bibfnamefont {J.~M.}\ \bibnamefont
  {Alameda}}, \bibinfo {author} {\bibfnamefont {D.}~\bibnamefont {Givord}},
  \bibinfo {author} {\bibfnamefont {R.}~\bibnamefont {Lemaire}}, \ and\
  \bibinfo {author} {\bibfnamefont {Q.}~\bibnamefont {Lu}},\ }\href {\doibase
  https://doi.org/10.1063/1.329622} {\bibfield  {journal} {\bibinfo  {journal}
  {J. Appl. Phys.}\ }\textbf {\bibinfo {volume} {52}},\ \bibinfo {pages} {2079}
  (\bibinfo {year} {1981})}\BibitemShut {NoStop}%
\bibitem [{\citenamefont {Klein}\ and\ \citenamefont
  {Menth}(1974)}]{Klein1974}%
  \BibitemOpen
  \bibfield  {author} {\bibinfo {author} {\bibfnamefont {H.~P.}\ \bibnamefont
  {Klein}}\ and\ \bibinfo {author} {\bibfnamefont {A.}~\bibnamefont {Menth}},\
  }\href {\doibase 10.1063/1.2947234} {\bibfield  {journal} {\bibinfo
  {journal} {AIP Conf. Proc.}\ }\textbf {\bibinfo {volume} {18}},\ \bibinfo
  {pages} {1177} (\bibinfo {year} {1974})}\BibitemShut {NoStop}%
\bibitem [{\citenamefont {Klein}\ \emph {et~al.}(1975)\citenamefont {Klein},
  \citenamefont {Menth},\ and\ \citenamefont {Perkins}}]{Klein1975}%
  \BibitemOpen
  \bibfield  {author} {\bibinfo {author} {\bibfnamefont {H.~P.}\ \bibnamefont
  {Klein}}, \bibinfo {author} {\bibfnamefont {A.}~\bibnamefont {Menth}}, \ and\
  \bibinfo {author} {\bibfnamefont {R.~S.}\ \bibnamefont {Perkins}},\ }\href
  {\doibase https://doi.org/10.1016/0378-4363(75)90061-3} {\bibfield  {journal}
  {\bibinfo  {journal} {Physica B+C}\ }\textbf {\bibinfo {volume} {80}},\
  \bibinfo {pages} {153} (\bibinfo {year} {1975})}\BibitemShut {NoStop}%
\bibitem [{\citenamefont {Zhao}\ \emph {et~al.}(1991)\citenamefont {Zhao},
  \citenamefont {Jin}, \citenamefont {Gr{\"{o}}ssinger}, \citenamefont {Kou},\
  and\ \citenamefont {Kirchmayr}}]{Zhao1991}%
  \BibitemOpen
  \bibfield  {author} {\bibinfo {author} {\bibfnamefont {T.~S.}\ \bibnamefont
  {Zhao}}, \bibinfo {author} {\bibfnamefont {H.~M.}\ \bibnamefont {Jin}},
  \bibinfo {author} {\bibfnamefont {R.}~\bibnamefont {Gr{\"{o}}ssinger}},
  \bibinfo {author} {\bibfnamefont {X.~C.}\ \bibnamefont {Kou}}, \ and\
  \bibinfo {author} {\bibfnamefont {H.~R.}\ \bibnamefont {Kirchmayr}},\ }\href
  {\doibase 10.1063/1.350020} {\bibfield  {journal} {\bibinfo  {journal} {J.
  Appl. Phys.}\ }\textbf {\bibinfo {volume} {70}},\ \bibinfo {pages} {6134}
  (\bibinfo {year} {1991})}\BibitemShut {NoStop}%
\bibitem [{\citenamefont {Skomski}(1998)}]{Skomski1998}%
  \BibitemOpen
  \bibfield  {author} {\bibinfo {author} {\bibfnamefont {R.}~\bibnamefont
  {Skomski}},\ }\href {\doibase https://doi.org/10.1063/1.367827} {\bibfield
  {journal} {\bibinfo  {journal} {J. Appl. Phys.}\ }\textbf {\bibinfo {volume}
  {83}},\ \bibinfo {pages} {6724} (\bibinfo {year} {1998})}\BibitemShut
  {NoStop}%
\bibitem [{\citenamefont {Nordstrom}\ \emph {et~al.}(1992)\citenamefont
  {Nordstrom}, \citenamefont {Brooks},\ and\ \citenamefont
  {Johansson}}]{Nordstorm1992}%
  \BibitemOpen
  \bibfield  {author} {\bibinfo {author} {\bibfnamefont {L.}~\bibnamefont
  {Nordstrom}}, \bibinfo {author} {\bibfnamefont {M.~S.~S.}\ \bibnamefont
  {Brooks}}, \ and\ \bibinfo {author} {\bibfnamefont {B.}~\bibnamefont
  {Johansson}},\ }\href {\doibase 10.1088/0953-8984/4/12/016} {\bibfield
  {journal} {\bibinfo  {journal} {J. Phys. Condens. Matter.}\ }\textbf
  {\bibinfo {volume} {4}},\ \bibinfo {pages} {3261} (\bibinfo {year}
  {1992})}\BibitemShut {NoStop}%
\bibitem [{\citenamefont {Daalderop}\ \emph {et~al.}(1996)\citenamefont
  {Daalderop}, \citenamefont {Kelly},\ and\ \citenamefont
  {Schuurmans}}]{Daalderop1996}%
  \BibitemOpen
  \bibfield  {author} {\bibinfo {author} {\bibfnamefont {G.}~\bibnamefont
  {Daalderop}}, \bibinfo {author} {\bibfnamefont {P.}~\bibnamefont {Kelly}}, \
  and\ \bibinfo {author} {\bibfnamefont {M.}~\bibnamefont {Schuurmans}},\
  }\href {\doibase 10.1103/PhysRevB.53.14415} {\bibfield  {journal} {\bibinfo
  {journal} {Phys. Rev. B}\ }\textbf {\bibinfo {volume} {53}},\ \bibinfo
  {pages} {14415} (\bibinfo {year} {1996})}\BibitemShut {NoStop}%
\bibitem [{\citenamefont {Yamaguchi}\ and\ \citenamefont
  {Asano}(1996)}]{Yamaguchi1996}%
  \BibitemOpen
  \bibfield  {author} {\bibinfo {author} {\bibfnamefont {M.}~\bibnamefont
  {Yamaguchi}}\ and\ \bibinfo {author} {\bibfnamefont {S.}~\bibnamefont
  {Asano}},\ }\href {\doibase 10.1063/1.362117} {\bibfield  {journal} {\bibinfo
   {journal} {J. Appl. Phys.}\ }\textbf {\bibinfo {volume} {79}},\ \bibinfo
  {pages} {5952} (\bibinfo {year} {1996})}\BibitemShut {NoStop}%
\bibitem [{\citenamefont {Zhu}\ \emph {et~al.}(2014)\citenamefont {Zhu},
  \citenamefont {Janoschek}, \citenamefont {Rosenberg}, \citenamefont
  {Ronning}, \citenamefont {Thompson}, \citenamefont {Torrez}, \citenamefont
  {Bauer},\ and\ \citenamefont {Batista}}]{Zhu2014}%
  \BibitemOpen
  \bibfield  {author} {\bibinfo {author} {\bibfnamefont {J.~X.}\ \bibnamefont
  {Zhu}}, \bibinfo {author} {\bibfnamefont {M.}~\bibnamefont {Janoschek}},
  \bibinfo {author} {\bibfnamefont {R.}~\bibnamefont {Rosenberg}}, \bibinfo
  {author} {\bibfnamefont {F.}~\bibnamefont {Ronning}}, \bibinfo {author}
  {\bibfnamefont {J.~D.}\ \bibnamefont {Thompson}}, \bibinfo {author}
  {\bibfnamefont {M.~A.}\ \bibnamefont {Torrez}}, \bibinfo {author}
  {\bibfnamefont {E.~D.}\ \bibnamefont {Bauer}}, \ and\ \bibinfo {author}
  {\bibfnamefont {C.~D.}\ \bibnamefont {Batista}},\ }\href
  {https://doi.org/10.1103/PhysRevX.4.021027} {\bibfield  {journal} {\bibinfo
  {journal} {Phys. Rev. X}\ }\textbf {\bibinfo {volume} {4}},\ \bibinfo {pages}
  {1} (\bibinfo {year} {2014})},\ \Eprint {http://arxiv.org/abs/1402.5543}
  {1402.5543} \BibitemShut {NoStop}%
\bibitem [{\citenamefont {Sakurai}\ \emph {et~al.}(2018)\citenamefont
  {Sakurai}, \citenamefont {Wu}, \citenamefont {Zhao}, \citenamefont {Nguyen},
  \citenamefont {Wang}, \citenamefont {Ho},\ and\ \citenamefont
  {Chelikowsky}}]{Sakurai2018}%
  \BibitemOpen
  \bibfield  {author} {\bibinfo {author} {\bibfnamefont {M.}~\bibnamefont
  {Sakurai}}, \bibinfo {author} {\bibfnamefont {S.}~\bibnamefont {Wu}},
  \bibinfo {author} {\bibfnamefont {X.}~\bibnamefont {Zhao}}, \bibinfo {author}
  {\bibfnamefont {M.~C.}\ \bibnamefont {Nguyen}}, \bibinfo {author}
  {\bibfnamefont {C.-Z.}\ \bibnamefont {Wang}}, \bibinfo {author}
  {\bibfnamefont {K.-M.}\ \bibnamefont {Ho}}, \ and\ \bibinfo {author}
  {\bibfnamefont {J.~R.}\ \bibnamefont {Chelikowsky}},\ }\href {\doibase
  https://doi.org/10.1103/PhysRevMaterials.2.084410} {\bibfield  {journal}
  {\bibinfo  {journal} {Phys. Rev. Mater.}\ }\textbf {\bibinfo {volume} {2}},\
  \bibinfo {pages} {084410} (\bibinfo {year} {2018})}\BibitemShut {NoStop}%
\bibitem [{\citenamefont {Nguyen}\ \emph {et~al.}(2018)\citenamefont {Nguyen},
  \citenamefont {Yao}, \citenamefont {Wang}, \citenamefont {Ho},\ and\
  \citenamefont {Antropov}}]{Nguyen2018}%
  \BibitemOpen
  \bibfield  {author} {\bibinfo {author} {\bibfnamefont {M.~C.}\ \bibnamefont
  {Nguyen}}, \bibinfo {author} {\bibfnamefont {Y.}~\bibnamefont {Yao}},
  \bibinfo {author} {\bibfnamefont {C.~Z.}\ \bibnamefont {Wang}}, \bibinfo
  {author} {\bibfnamefont {K.~M.}\ \bibnamefont {Ho}}, \ and\ \bibinfo {author}
  {\bibfnamefont {V.~P.}\ \bibnamefont {Antropov}},\ }\href
  {https://doi.org/10.1088/1361-648X/aab9fa} {\bibfield  {journal} {\bibinfo
  {journal} {J. Phys. Condens. Matter.}\ }\textbf {\bibinfo {volume} {30}}
  (\bibinfo {year} {2018})}\BibitemShut {NoStop}%
\bibitem [{\citenamefont {Matsumoto}\ \emph {et~al.}(2015)\citenamefont
  {Matsumoto}, \citenamefont {Banerjee},\ and\ \citenamefont
  {Staunton}}]{Matsumoto2015}%
  \BibitemOpen
  \bibfield  {author} {\bibinfo {author} {\bibfnamefont {M.}~\bibnamefont
  {Matsumoto}}, \bibinfo {author} {\bibfnamefont {R.}~\bibnamefont {Banerjee}},
  \ and\ \bibinfo {author} {\bibfnamefont {J.~B.}\ \bibnamefont {Staunton}},\
  }\href {\doibase 10.7566/JPSCP.5.011004} {\bibfield  {journal} {\bibinfo
  {journal} {Phys. Soc. Jpn. Conf. Proc.}\ }\textbf {\bibinfo {volume} {5}},\
  \bibinfo {pages} {011004} (\bibinfo {year} {2015})}\BibitemShut {NoStop}%
\bibitem [{\citenamefont {Steinbeck}\ \emph {et~al.}(2001)\citenamefont
  {Steinbeck}, \citenamefont {Richter},\ and\ \citenamefont
  {Eschrig}}]{Steinbeck2001}%
  \BibitemOpen
  \bibfield  {author} {\bibinfo {author} {\bibfnamefont {L.}~\bibnamefont
  {Steinbeck}}, \bibinfo {author} {\bibfnamefont {M.}~\bibnamefont {Richter}},
  \ and\ \bibinfo {author} {\bibfnamefont {H.}~\bibnamefont {Eschrig}},\ }\href
  {\doibase 10.1103/PhysRevB.63.184431} {\bibfield  {journal} {\bibinfo
  {journal} {Phys. Rev. B}\ }\textbf {\bibinfo {volume} {63}},\ \bibinfo
  {pages} {22} (\bibinfo {year} {2001})}\BibitemShut {NoStop}%
\bibitem [{\citenamefont {Larson}\ and\ \citenamefont
  {Mazin}(2003)}]{Larson2003}%
  \BibitemOpen
  \bibfield  {author} {\bibinfo {author} {\bibfnamefont {P.}~\bibnamefont
  {Larson}}\ and\ \bibinfo {author} {\bibfnamefont {I.~I.}\ \bibnamefont
  {Mazin}},\ }\href {\doibase 10.1063/1.1556154} {\bibfield  {journal}
  {\bibinfo  {journal} {J. Appl. Phys.}\ }\textbf {\bibinfo {volume} {93}},\
  \bibinfo {pages} {6888} (\bibinfo {year} {2003})}\BibitemShut {NoStop}%
\bibitem [{\citenamefont {Larson}\ \emph {et~al.}(2004)\citenamefont {Larson},
  \citenamefont {Mazin},\ and\ \citenamefont
  {Papaconstantopoulos}}]{Larson2004}%
  \BibitemOpen
  \bibfield  {author} {\bibinfo {author} {\bibfnamefont {P.}~\bibnamefont
  {Larson}}, \bibinfo {author} {\bibfnamefont {I.~I.}\ \bibnamefont {Mazin}}, \
  and\ \bibinfo {author} {\bibfnamefont {D.~A.}\ \bibnamefont
  {Papaconstantopoulos}},\ }\href {\doibase 10.1103/PhysRevB.69.134408}
  {\bibfield  {journal} {\bibinfo  {journal} {Phys. Rev. B}\ }\textbf {\bibinfo
  {volume} {69}},\ \bibinfo {pages} {2} (\bibinfo {year} {2004})}\BibitemShut
  {NoStop}%
\bibitem [{\citenamefont {Liu}\ \emph {et~al.}(2010)\citenamefont {Liu},
  \citenamefont {Altounian},\ and\ \citenamefont {Yue}}]{Liu2010}%
  \BibitemOpen
  \bibfield  {author} {\bibinfo {author} {\bibfnamefont {X.~B.}\ \bibnamefont
  {Liu}}, \bibinfo {author} {\bibfnamefont {Z.}~\bibnamefont {Altounian}}, \
  and\ \bibinfo {author} {\bibfnamefont {M.}~\bibnamefont {Yue}},\ }\href
  {https://doi.org/10.1063/1.3339020} {\bibfield  {journal} {\bibinfo
  {journal} {J. Appl. Phys.}\ }\textbf {\bibinfo {volume} {107}} (\bibinfo
  {year} {2010})}\BibitemShut {NoStop}%
\bibitem [{\citenamefont {Patrick}\ \emph {et~al.}(2017)\citenamefont
  {Patrick}, \citenamefont {Kumar}, \citenamefont {Balakrishnan}, \citenamefont
  {Edwards}, \citenamefont {Lees}, \citenamefont {Mendive-Tapia}, \citenamefont
  {Petit},\ and\ \citenamefont {Staunton}}]{Patrick2017}%
  \BibitemOpen
  \bibfield  {author} {\bibinfo {author} {\bibfnamefont {C.~E.}\ \bibnamefont
  {Patrick}}, \bibinfo {author} {\bibfnamefont {S.}~\bibnamefont {Kumar}},
  \bibinfo {author} {\bibfnamefont {G.}~\bibnamefont {Balakrishnan}}, \bibinfo
  {author} {\bibfnamefont {R.~S.}\ \bibnamefont {Edwards}}, \bibinfo {author}
  {\bibfnamefont {M.~R.}\ \bibnamefont {Lees}}, \bibinfo {author}
  {\bibfnamefont {E.}~\bibnamefont {Mendive-Tapia}}, \bibinfo {author}
  {\bibfnamefont {L.}~\bibnamefont {Petit}}, \ and\ \bibinfo {author}
  {\bibfnamefont {J.~B.}\ \bibnamefont {Staunton}},\ }\href {\doibase
  https://doi.org/10.1103/PhysRevMaterials.1.024411} {\bibfield  {journal}
  {\bibinfo  {journal} {Phys. Rev. Mater.}\ }\textbf {\bibinfo {volume} {1}},\
  \bibinfo {pages} {1} (\bibinfo {year} {2017})}\BibitemShut {NoStop}%
\bibitem [{\citenamefont {Patrick}\ \emph {et~al.}(2019)\citenamefont
  {Patrick}, \citenamefont {Matsumoto},\ and\ \citenamefont
  {Staunton}}]{Patrick2019}%
  \BibitemOpen
  \bibfield  {author} {\bibinfo {author} {\bibfnamefont {C.~E.}\ \bibnamefont
  {Patrick}}, \bibinfo {author} {\bibfnamefont {M.}~\bibnamefont {Matsumoto}},
  \ and\ \bibinfo {author} {\bibfnamefont {J.~B.}\ \bibnamefont {Staunton}},\
  }\href {\doibase 10.1016/j.jmmm.2019.01.061} {\bibfield  {journal} {\bibinfo
  {journal} {J. Magn. Magn. Mater.}\ }\textbf {\bibinfo {volume} {477}},\
  \bibinfo {pages} {147} (\bibinfo {year} {2019})}\BibitemShut {NoStop}%
\bibitem [{\citenamefont {Asali}\ \emph {et~al.}(2019)\citenamefont {Asali},
  \citenamefont {Fidler},\ and\ \citenamefont {Suess}}]{Asali2019}%
  \BibitemOpen
  \bibfield  {author} {\bibinfo {author} {\bibfnamefont {A.}~\bibnamefont
  {Asali}}, \bibinfo {author} {\bibfnamefont {J.}~\bibnamefont {Fidler}}, \
  and\ \bibinfo {author} {\bibfnamefont {D.}~\bibnamefont {Suess}},\ }\href
  {\doibase 10.1016/j.jmmm.2019.04.047} {\bibfield  {journal} {\bibinfo
  {journal} {J. Magn. Magn. Mater.}\ }\textbf {\bibinfo {volume} {485}},\
  \bibinfo {pages} {61} (\bibinfo {year} {2019})}\BibitemShut {NoStop}%
\bibitem [{\citenamefont {Crisan}\ \emph {et~al.}(1995)\citenamefont {Crisan},
  \citenamefont {Popescu}, \citenamefont {Vernes}, \citenamefont {Andreica},
  \citenamefont {Burda},\ and\ \citenamefont {Cristea}}]{Crisan1995}%
  \BibitemOpen
  \bibfield  {author} {\bibinfo {author} {\bibfnamefont {V.}~\bibnamefont
  {Crisan}}, \bibinfo {author} {\bibfnamefont {V.}~\bibnamefont {Popescu}},
  \bibinfo {author} {\bibfnamefont {A.}~\bibnamefont {Vernes}}, \bibinfo
  {author} {\bibfnamefont {D.}~\bibnamefont {Andreica}}, \bibinfo {author}
  {\bibfnamefont {I.}~\bibnamefont {Burda}}, \ and\ \bibinfo {author}
  {\bibfnamefont {S.}~\bibnamefont {Cristea}},\ }\href {\doibase
  https://doi.org/10.1016/0925-8388(94)01494-9} {\bibfield  {journal} {\bibinfo
   {journal} {J. Alloys Compd}\ }\textbf {\bibinfo {volume} {223}},\ \bibinfo
  {pages} {147} (\bibinfo {year} {1995})}\BibitemShut {NoStop}%
\bibitem [{\citenamefont {Yamada}\ \emph {et~al.}(1999)\citenamefont {Yamada},
  \citenamefont {Terao}, \citenamefont {Morozumi}, \citenamefont {Terao},\ and\
  \citenamefont {Yamada}}]{Yamada1999}%
  \BibitemOpen
  \bibfield  {author} {\bibinfo {author} {\bibfnamefont {H.}~\bibnamefont
  {Yamada}}, \bibinfo {author} {\bibfnamefont {K.}~\bibnamefont {Terao}},
  \bibinfo {author} {\bibfnamefont {H.}~\bibnamefont {Morozumi}}, \bibinfo
  {author} {\bibfnamefont {K.}~\bibnamefont {Terao}}, \ and\ \bibinfo {author}
  {\bibfnamefont {H.}~\bibnamefont {Yamada}},\ }\href {\doibase
  https://doi.org/10.1088/0953-8984/11/2/013} {\bibfield  {journal} {\bibinfo
  {journal} {J. Phys. Condens. Matter.}\ }\textbf {\bibinfo {volume} {11}},\
  \bibinfo {pages} {483} (\bibinfo {year} {1999})}\BibitemShut {NoStop}%
\bibitem [{\citenamefont {Ishikawa}\ \emph {et~al.}(2003)\citenamefont
  {Ishikawa}, \citenamefont {Yamamoto}, \citenamefont {Umehara}, \citenamefont
  {Yamaguchi}, \citenamefont {Bartashevich}, \citenamefont {Mitamura},
  \citenamefont {Goto},\ and\ \citenamefont {Yamada}}]{Ishikawa2003}%
  \BibitemOpen
  \bibfield  {author} {\bibinfo {author} {\bibfnamefont {F.}~\bibnamefont
  {Ishikawa}}, \bibinfo {author} {\bibfnamefont {I.}~\bibnamefont {Yamamoto}},
  \bibinfo {author} {\bibfnamefont {I.}~\bibnamefont {Umehara}}, \bibinfo
  {author} {\bibfnamefont {M.}~\bibnamefont {Yamaguchi}}, \bibinfo {author}
  {\bibfnamefont {M.~I.}\ \bibnamefont {Bartashevich}}, \bibinfo {author}
  {\bibfnamefont {H.}~\bibnamefont {Mitamura}}, \bibinfo {author}
  {\bibfnamefont {T.}~\bibnamefont {Goto}}, \ and\ \bibinfo {author}
  {\bibfnamefont {H.}~\bibnamefont {Yamada}},\ }\href {\doibase
  https://doi.org/10.1016/S0921-4526(02)02639-X} {\bibfield  {journal}
  {\bibinfo  {journal} {Physica B: Condens. Matter.}\ }\textbf {\bibinfo
  {volume} {328}},\ \bibinfo {pages} {386} (\bibinfo {year}
  {2003})}\BibitemShut {NoStop}%
\bibitem [{\citenamefont {Landa}\ \emph {et~al.}(2020)\citenamefont {Landa},
  \citenamefont {S{\"{o}}derlind}, \citenamefont {Moore},\ and\ \citenamefont
  {Perron}}]{Landa2020}%
  \BibitemOpen
  \bibfield  {author} {\bibinfo {author} {\bibfnamefont {A.}~\bibnamefont
  {Landa}}, \bibinfo {author} {\bibfnamefont {P.}~\bibnamefont
  {S{\"{o}}derlind}}, \bibinfo {author} {\bibfnamefont {E.~E.}\ \bibnamefont
  {Moore}}, \ and\ \bibinfo {author} {\bibfnamefont {A.}~\bibnamefont
  {Perron}},\ }\href {\doibase 10.3390/app10176037} {\bibfield  {journal}
  {\bibinfo  {journal} {Appl. Sci. (Switz)}\ }\textbf {\bibinfo {volume}
  {10}},\ \bibinfo {pages} {1} (\bibinfo {year} {2020})}\BibitemShut {NoStop}%
\bibitem [{\citenamefont {Ogura}\ and\ \citenamefont {Akai}(2005)}]{Ogura2005}%
  \BibitemOpen
  \bibfield  {author} {\bibinfo {author} {\bibfnamefont {M.}~\bibnamefont
  {Ogura}}\ and\ \bibinfo {author} {\bibfnamefont {H.}~\bibnamefont {Akai}},\
  }\href {\doibase 10.1088/0953-8984/17/37/011} {\bibfield  {journal} {\bibinfo
   {journal} {J. Phys. Condens. Matter.}\ }\textbf {\bibinfo {volume} {17}},\
  \bibinfo {pages} {5741} (\bibinfo {year} {2005})}\BibitemShut {NoStop}%
\bibitem [{\citenamefont {Shiba}(1971)}]{Shiba1971}%
  \BibitemOpen
  \bibfield  {author} {\bibinfo {author} {\bibfnamefont {H.}~\bibnamefont
  {Shiba}},\ }\href {\doibase 10.1143/PTP.46.77} {\bibfield  {journal}
  {\bibinfo  {journal} {Prog. Theor. Phys.}\ }\textbf {\bibinfo {volume}
  {46}},\ \bibinfo {pages} {77} (\bibinfo {year} {1971})}\BibitemShut {NoStop}%
\bibitem [{\citenamefont {Soven}(1970)}]{Soven1970}%
  \BibitemOpen
  \bibfield  {author} {\bibinfo {author} {\bibfnamefont {P.}~\bibnamefont
  {Soven}},\ }\href {\doibase 10.1103/PhysRevB.2.4715} {\bibfield  {journal}
  {\bibinfo  {journal} {Phys. Rev. B}\ }\textbf {\bibinfo {volume} {2}},\
  \bibinfo {pages} {4715} (\bibinfo {year} {1970})}\BibitemShut {NoStop}%
\bibitem [{\citenamefont {Moruzzi}\ \emph {et~al.}(1978)\citenamefont
  {Moruzzi}, \citenamefont {Janak},\ and\ \citenamefont
  {Williams}}]{Moruzzi1978}%
  \BibitemOpen
  \bibfield  {author} {\bibinfo {author} {\bibfnamefont {V.~L.}\ \bibnamefont
  {Moruzzi}}, \bibinfo {author} {\bibfnamefont {J.~F.}\ \bibnamefont {Janak}},
  \ and\ \bibinfo {author} {\bibfnamefont {A.~R.}\ \bibnamefont {Williams}},\
  }in\ \href {\doibase https://doi.org/10.1016/B978-0-08-022705-4.50004-1}
  {\emph {\bibinfo {booktitle} {{Calculated Electronic Properties of
  Metals}}}},\ \bibinfo {editor} {edited by\ \bibinfo {editor} {\bibfnamefont
  {V.~L.}\ \bibnamefont {Moruzzi}}, \bibinfo {editor} {\bibfnamefont {J.~F.}\
  \bibnamefont {Janak}}, \ and\ \bibinfo {editor} {\bibfnamefont {A.~R.}\
  \bibnamefont {Williams}}}\ (\bibinfo  {publisher} {Pergamon},\ \bibinfo
  {year} {1978})\BibitemShut {NoStop}%
\bibitem [{\citenamefont {Perdew}\ \emph {et~al.}(1996)\citenamefont {Perdew},
  \citenamefont {Burke},\ and\ \citenamefont {Ernzerhof}}]{Perdew1996}%
  \BibitemOpen
  \bibfield  {author} {\bibinfo {author} {\bibfnamefont {J.~P.}\ \bibnamefont
  {Perdew}}, \bibinfo {author} {\bibfnamefont {K.}~\bibnamefont {Burke}}, \
  and\ \bibinfo {author} {\bibfnamefont {M.}~\bibnamefont {Ernzerhof}},\ }\href
  {\doibase 10.1103/PhysRevLett.77.3865} {\bibfield  {journal} {\bibinfo
  {journal} {Phys. Rev. Lett.}\ }\textbf {\bibinfo {volume} {77}},\ \bibinfo
  {pages} {3865} (\bibinfo {year} {1996})},\ \bibinfo {note} {erratum: Phys.
  Rev. Lett.. 1997 Feb;78:1396^^e2^^80^^931396.}\BibitemShut {Stop}%
\bibitem [{\citenamefont {Heidemann}\ \emph {et~al.}(1975)\citenamefont
  {Heidemann}, \citenamefont {Richter},\ and\ \citenamefont
  {Buschow}}]{Heidemann1975}%
  \BibitemOpen
  \bibfield  {author} {\bibinfo {author} {\bibfnamefont {A.}~\bibnamefont
  {Heidemann}}, \bibinfo {author} {\bibfnamefont {D.}~\bibnamefont {Richter}},
  \ and\ \bibinfo {author} {\bibfnamefont {K.~H.}\ \bibnamefont {Buschow}},\
  }\href {\doibase 10.1007/BF01312807} {\bibfield  {journal} {\bibinfo
  {journal} {Z. Phys. B}\ }\textbf {\bibinfo {volume} {22}},\ \bibinfo {pages}
  {367} (\bibinfo {year} {1975})}\BibitemShut {NoStop}%
\bibitem [{\citenamefont {Schweizer}\ and\ \citenamefont
  {Tasset}(1980)}]{Schweizer1980}%
  \BibitemOpen
  \bibfield  {author} {\bibinfo {author} {\bibfnamefont {J.}~\bibnamefont
  {Schweizer}}\ and\ \bibinfo {author} {\bibfnamefont {F.}~\bibnamefont
  {Tasset}},\ }\href {\doibase 10.1088/0305-4608/10/12/020} {\bibfield
  {journal} {\bibinfo  {journal} {J. Phys. F. Met. Phys.}\ }\textbf {\bibinfo
  {volume} {10}},\ \bibinfo {pages} {2799} (\bibinfo {year}
  {1980})}\BibitemShut {NoStop}%
\bibitem [{\citenamefont {Chuang}\ \emph {et~al.}(1982)\citenamefont {Chuang},
  \citenamefont {Wu},\ and\ \citenamefont {Chang}}]{Chuang1982}%
  \BibitemOpen
  \bibfield  {author} {\bibinfo {author} {\bibfnamefont {Y.~C.}\ \bibnamefont
  {Chuang}}, \bibinfo {author} {\bibfnamefont {C.~H.}\ \bibnamefont {Wu}}, \
  and\ \bibinfo {author} {\bibfnamefont {Y.~C.}\ \bibnamefont {Chang}},\ }\href
  {\doibase https://doi.org/10.1016/0022-5088(82)90145-X} {\bibfield  {journal}
  {\bibinfo  {journal} {J. Less. Common. Met.}\ }\textbf {\bibinfo {volume}
  {84}},\ \bibinfo {pages} {201} (\bibinfo {year} {1982})}\BibitemShut
  {NoStop}%
\bibitem [{\citenamefont {Deportes}\ \emph {et~al.}(1976)\citenamefont
  {Deportes}, \citenamefont {Givord}, \citenamefont {Schweizer},\ and\
  \citenamefont {Tasset}}]{Deportes1976}%
  \BibitemOpen
  \bibfield  {author} {\bibinfo {author} {\bibfnamefont {J.}~\bibnamefont
  {Deportes}}, \bibinfo {author} {\bibfnamefont {D.}~\bibnamefont {Givord}},
  \bibinfo {author} {\bibfnamefont {J.}~\bibnamefont {Schweizer}}, \ and\
  \bibinfo {author} {\bibfnamefont {F.}~\bibnamefont {Tasset}},\ }\href
  {\doibase 10.1109/TMAG.1976.1059185} {\bibfield  {journal} {\bibinfo
  {journal} {IEEE Trans. Magn.}\ }\textbf {\bibinfo {volume} {12}},\ \bibinfo
  {pages} {1000} (\bibinfo {year} {1976})}\BibitemShut {NoStop}%
\bibitem [{\citenamefont {Kresse}\ and\ \citenamefont
  {Furthm{\"u}ller}(1996)}]{Kresse1996}%
  \BibitemOpen
  \bibfield  {author} {\bibinfo {author} {\bibfnamefont {G.}~\bibnamefont
  {Kresse}}\ and\ \bibinfo {author} {\bibfnamefont {J.}~\bibnamefont
  {Furthm{\"u}ller}},\ }\href {\doibase 10.1016/0927-0256(96)00008-0}
  {\bibfield  {journal} {\bibinfo  {journal} {Comput. Mater. Sci.}\ }\textbf
  {\bibinfo {volume} {6}},\ \bibinfo {pages} {15} (\bibinfo {year}
  {1996})}\BibitemShut {NoStop}%
\bibitem [{\citenamefont {Maruyama}\ \emph {et~al.}(1999)\citenamefont
  {Maruyama}, \citenamefont {Nagai}, \citenamefont {Amako}, \citenamefont
  {Yoshie},\ and\ \citenamefont {Adachi}}]{Maruyama1999}%
  \BibitemOpen
  \bibfield  {author} {\bibinfo {author} {\bibfnamefont {F.}~\bibnamefont
  {Maruyama}}, \bibinfo {author} {\bibfnamefont {H.}~\bibnamefont {Nagai}},
  \bibinfo {author} {\bibfnamefont {Y.}~\bibnamefont {Amako}}, \bibinfo
  {author} {\bibfnamefont {H.}~\bibnamefont {Yoshie}}, \ and\ \bibinfo {author}
  {\bibfnamefont {K.}~\bibnamefont {Adachi}},\ }\href {\doibase
  https://doi.org/10.1016/S0921-4526(98)01450-1} {\bibfield  {journal}
  {\bibinfo  {journal} {Physica B: Condens. Matter.}\ }\textbf {\bibinfo
  {volume} {266}},\ \bibinfo {pages} {356} (\bibinfo {year}
  {1999})}\BibitemShut {NoStop}%
\bibitem [{\citenamefont {Rosner}\ \emph {et~al.}(2006)\citenamefont {Rosner},
  \citenamefont {Koudela}, \citenamefont {Schwarz}, \citenamefont {Handstein},
  \citenamefont {Hanfland}, \citenamefont {Opahle}, \citenamefont {Koepernik},
  \citenamefont {Kuz'min}, \citenamefont {M{\"{u}}ller}, \citenamefont
  {Mydosh},\ and\ \citenamefont {Richter}}]{Rosner2006}%
  \BibitemOpen
  \bibfield  {author} {\bibinfo {author} {\bibfnamefont {H.}~\bibnamefont
  {Rosner}}, \bibinfo {author} {\bibfnamefont {D.}~\bibnamefont {Koudela}},
  \bibinfo {author} {\bibfnamefont {U.}~\bibnamefont {Schwarz}}, \bibinfo
  {author} {\bibfnamefont {A.}~\bibnamefont {Handstein}}, \bibinfo {author}
  {\bibfnamefont {M.}~\bibnamefont {Hanfland}}, \bibinfo {author}
  {\bibfnamefont {I.}~\bibnamefont {Opahle}}, \bibinfo {author} {\bibfnamefont
  {K.}~\bibnamefont {Koepernik}}, \bibinfo {author} {\bibfnamefont {M.~D.}\
  \bibnamefont {Kuz'min}}, \bibinfo {author} {\bibfnamefont {K.~H.}\
  \bibnamefont {M{\"{u}}ller}}, \bibinfo {author} {\bibfnamefont {J.~A.}\
  \bibnamefont {Mydosh}}, \ and\ \bibinfo {author} {\bibfnamefont
  {M.}~\bibnamefont {Richter}},\ }\href {\doibase 10.1038/nphys341} {\bibfield
  {journal} {\bibinfo  {journal} {Nat. Phys.}\ }\textbf {\bibinfo {volume}
  {2}},\ \bibinfo {pages} {469} (\bibinfo {year} {2006})}\BibitemShut {NoStop}%
\bibitem [{\citenamefont {Frederick}\ and\ \citenamefont
  {Hoch}(1974)}]{Frederick1974}%
  \BibitemOpen
  \bibfield  {author} {\bibinfo {author} {\bibfnamefont {W.}~\bibnamefont
  {Frederick}}\ and\ \bibinfo {author} {\bibfnamefont {M.}~\bibnamefont
  {Hoch}},\ }\href {\doibase 10.1109/TMAG.1974.1058379} {\bibfield  {journal}
  {\bibinfo  {journal} {IEEE Trans. Magn.}\ }\textbf {\bibinfo {volume} {10}},\
  \bibinfo {pages} {733} (\bibinfo {year} {1974})}\BibitemShut {NoStop}%
\bibitem [{\citenamefont {Ucar}\ \emph {et~al.}(2020)\citenamefont {Ucar},
  \citenamefont {Choudhary},\ and\ \citenamefont {Paudyal}}]{Ucar2020}%
  \BibitemOpen
  \bibfield  {author} {\bibinfo {author} {\bibfnamefont {H.}~\bibnamefont
  {Ucar}}, \bibinfo {author} {\bibfnamefont {R.}~\bibnamefont {Choudhary}}, \
  and\ \bibinfo {author} {\bibfnamefont {D.}~\bibnamefont {Paudyal}},\ }\href
  {\doibase https://doi.org/10.1016/j.jmmm.2019.165902} {\bibfield  {journal}
  {\bibinfo  {journal} {J. Magn. Magn. Mater.}\ }\textbf {\bibinfo {volume}
  {496}},\ \bibinfo {pages} {165902} (\bibinfo {year} {2020})}\BibitemShut
  {NoStop}%
\bibitem [{\citenamefont {Plugaru}\ \emph {et~al.}(2014)\citenamefont
  {Plugaru}, \citenamefont {Valeanu}, \citenamefont {Plugaru},\ and\
  \citenamefont {Campo}}]{Plugaru2014}%
  \BibitemOpen
  \bibfield  {author} {\bibinfo {author} {\bibfnamefont {N.}~\bibnamefont
  {Plugaru}}, \bibinfo {author} {\bibfnamefont {M.}~\bibnamefont {Valeanu}},
  \bibinfo {author} {\bibfnamefont {R.}~\bibnamefont {Plugaru}}, \ and\
  \bibinfo {author} {\bibfnamefont {J.}~\bibnamefont {Campo}},\ }\href
  {\doibase 10.1063/1.4862163} {\bibfield  {journal} {\bibinfo  {journal} {J.
  Appl. Phys.}\ }\textbf {\bibinfo {volume} {115}},\ \bibinfo {pages} {23907}
  (\bibinfo {year} {2014})}\BibitemShut {NoStop}%
\bibitem [{\citenamefont {Burzo}\ \emph {et~al.}(2020)\citenamefont {Burzo},
  \citenamefont {Vlaic}, \citenamefont {Kozlenko}, \citenamefont {Golosova},
  \citenamefont {Kichanov}, \citenamefont {Savenko}, \citenamefont {Ostlin},\
  and\ \citenamefont {Chioncel}}]{Burzo2020}%
  \BibitemOpen
  \bibfield  {author} {\bibinfo {author} {\bibfnamefont {E.}~\bibnamefont
  {Burzo}}, \bibinfo {author} {\bibfnamefont {P.}~\bibnamefont {Vlaic}},
  \bibinfo {author} {\bibfnamefont {D.~P.}\ \bibnamefont {Kozlenko}}, \bibinfo
  {author} {\bibfnamefont {N.~O.}\ \bibnamefont {Golosova}}, \bibinfo {author}
  {\bibfnamefont {S.~E.}\ \bibnamefont {Kichanov}}, \bibinfo {author}
  {\bibfnamefont {B.~N.}\ \bibnamefont {Savenko}}, \bibinfo {author}
  {\bibfnamefont {A.}~\bibnamefont {Ostlin}}, \ and\ \bibinfo {author}
  {\bibfnamefont {L.}~\bibnamefont {Chioncel}},\ }\href {\doibase
  https://doi.org/10.1016/j.jmst.2019.12.001} {\bibfield  {journal} {\bibinfo
  {journal} {J. Mater. Sci. Technol.}\ }\textbf {\bibinfo {volume} {42}},\
  \bibinfo {pages} {106} (\bibinfo {year} {2020})}\BibitemShut {NoStop}%
\bibitem [{\citenamefont {Inomata}(1981)}]{Inomata1981}%
  \BibitemOpen
  \bibfield  {author} {\bibinfo {author} {\bibfnamefont {K.}~\bibnamefont
  {Inomata}},\ }\href {\doibase 10.1103/PhysRevB.23.2076} {\bibfield  {journal}
  {\bibinfo  {journal} {Phys. Rev. B}\ }\textbf {\bibinfo {volume} {23}},\
  \bibinfo {pages} {2076} (\bibinfo {year} {1981})}\BibitemShut {NoStop}%
\bibitem [{\citenamefont {Franse}\ \emph {et~al.}(1988)\citenamefont {Franse},
  \citenamefont {Thuy},\ and\ \citenamefont {Hong}}]{Franse1988}%
  \BibitemOpen
  \bibfield  {author} {\bibinfo {author} {\bibfnamefont {J.~J.~M.}\
  \bibnamefont {Franse}}, \bibinfo {author} {\bibfnamefont {N.~P.}\
  \bibnamefont {Thuy}}, \ and\ \bibinfo {author} {\bibfnamefont {N.~M.}\
  \bibnamefont {Hong}},\ }\href {https://doi.org/10.1016/0304-8853(88)90235-1}
  {\bibfield  {journal} {\bibinfo  {journal} {J. Magn. Magn. Mater.}\ }\textbf
  {\bibinfo {volume} {72}},\ \bibinfo {pages} {361} (\bibinfo {year}
  {1988})}\BibitemShut {NoStop}%
\bibitem [{\citenamefont {Tellez-Blanco}\ \emph {et~al.}(2000)\citenamefont
  {Tellez-Blanco}, \citenamefont {Grossinger}, \citenamefont {Turtelli},\ and\
  \citenamefont {Estevez-Rams}}]{Tellez2000}%
  \BibitemOpen
  \bibfield  {author} {\bibinfo {author} {\bibfnamefont {J.~C.}\ \bibnamefont
  {Tellez-Blanco}}, \bibinfo {author} {\bibfnamefont {R.}~\bibnamefont
  {Grossinger}}, \bibinfo {author} {\bibfnamefont {R.~S.}\ \bibnamefont
  {Turtelli}}, \ and\ \bibinfo {author} {\bibfnamefont {E.}~\bibnamefont
  {Estevez-Rams}},\ }\href {\doibase 10.1109/20.908790} {\bibfield  {journal}
  {\bibinfo  {journal} {IEEE Trans. Magn.}\ }\textbf {\bibinfo {volume} {36}},\
  \bibinfo {pages} {3333} (\bibinfo {year} {2000})}\BibitemShut {NoStop}%
\bibitem [{\citenamefont {Tedstone}\ \emph {et~al.}(2019)\citenamefont
  {Tedstone}, \citenamefont {Patrick}, \citenamefont {Kumar}, \citenamefont
  {Edwards}, \citenamefont {Lees}, \citenamefont {Balakrishnan},\ and\
  \citenamefont {Staunton}}]{Tedstone2019}%
  \BibitemOpen
  \bibfield  {author} {\bibinfo {author} {\bibfnamefont {A.~L.}\ \bibnamefont
  {Tedstone}}, \bibinfo {author} {\bibfnamefont {C.~E.}\ \bibnamefont
  {Patrick}}, \bibinfo {author} {\bibfnamefont {S.}~\bibnamefont {Kumar}},
  \bibinfo {author} {\bibfnamefont {R.~S.}\ \bibnamefont {Edwards}}, \bibinfo
  {author} {\bibfnamefont {M.~R.}\ \bibnamefont {Lees}}, \bibinfo {author}
  {\bibfnamefont {G.}~\bibnamefont {Balakrishnan}}, \ and\ \bibinfo {author}
  {\bibfnamefont {J.~B.}\ \bibnamefont {Staunton}},\ }\href {\doibase
  https://doi.org/10.1103/PhysRevMaterials.3.034409} {\bibfield  {journal}
  {\bibinfo  {journal} {Phys. Rev. Mater.}\ }\textbf {\bibinfo {volume} {3}},\
  \bibinfo {pages} {1} (\bibinfo {year} {2019})}\BibitemShut {NoStop}%
\bibitem [{\citenamefont {Buschow}\ and\ \citenamefont
  {Brouha}(1976)}]{Buschow1976}%
  \BibitemOpen
  \bibfield  {author} {\bibinfo {author} {\bibfnamefont {K.~H.~J.}\
  \bibnamefont {Buschow}}\ and\ \bibinfo {author} {\bibfnamefont
  {M.}~\bibnamefont {Brouha}},\ }\href {\doibase 10.1063/1.30485} {\bibfield
  {journal} {\bibinfo  {journal} {AIP Conf. Proc.}\ }\textbf {\bibinfo {volume}
  {29}},\ \bibinfo {pages} {618} (\bibinfo {year} {1976})}\BibitemShut
  {NoStop}%
\bibitem [{sup()}]{suppl}%
  \BibitemOpen
  \href@noop {} {\ }\bibinfo {note} {Supplemental Material are not available
  now.}\BibitemShut {Stop}%
\bibitem [{\citenamefont {S{\"{o}}derlind}\ \emph {et~al.}(2017)\citenamefont
  {S{\"{o}}derlind}, \citenamefont {Landa}, \citenamefont {Locht},
  \citenamefont {{\AA}berg}, \citenamefont {Kvashnin}, \citenamefont {Pereiro},
  \citenamefont {D{\"{a}}ne}, \citenamefont {Turchi}, \citenamefont
  {Antropov},\ and\ \citenamefont {Eriksson}}]{Soderlind2017}%
  \BibitemOpen
  \bibfield  {author} {\bibinfo {author} {\bibfnamefont {P.}~\bibnamefont
  {S{\"{o}}derlind}}, \bibinfo {author} {\bibfnamefont {A.}~\bibnamefont
  {Landa}}, \bibinfo {author} {\bibfnamefont {I.~L.}\ \bibnamefont {Locht}},
  \bibinfo {author} {\bibfnamefont {D.}~\bibnamefont {{\AA}berg}}, \bibinfo
  {author} {\bibfnamefont {Y.}~\bibnamefont {Kvashnin}}, \bibinfo {author}
  {\bibfnamefont {M.}~\bibnamefont {Pereiro}}, \bibinfo {author} {\bibfnamefont
  {M.}~\bibnamefont {D{\"{a}}ne}}, \bibinfo {author} {\bibfnamefont {P.~E.}\
  \bibnamefont {Turchi}}, \bibinfo {author} {\bibfnamefont {V.~P.}\
  \bibnamefont {Antropov}}, \ and\ \bibinfo {author} {\bibfnamefont
  {O.}~\bibnamefont {Eriksson}},\ }\href {\doibase 10.1103/PhysRevB.96.100404}
  {\bibfield  {journal} {\bibinfo  {journal} {Phys. Rev. B}\ }\textbf {\bibinfo
  {volume} {96}},\ \bibinfo {pages} {1} (\bibinfo {year} {2017})}\BibitemShut
  {NoStop}%
\end{thebibliography}%
\bibliographystyle{apsrev4-1}

\end{document}